\documentclass[aoas]{imsart}

\RequirePackage{amsthm,amsmath,amsfonts,amssymb}
\RequirePackage[authoryear]{natbib}
\RequirePackage[colorlinks,citecolor=blue,urlcolor=blue]{hyperref}
\RequirePackage{graphicx}
\RequirePackage{amsmath, amssymb, multirow, graphicx,footnote,caption,color,bm}

\startlocaldefs

\numberwithin{equation}{section}

\endlocaldefs

\begin{document}

\begin{frontmatter}
\title{A Semi-parametric Promotion Time Cure Model with Support Vector Machine}
\runtitle{Promotion Time Cure Model with Support Vector Machine}

\begin{aug}
\author[A]{\fnms{Suvra}~\snm{Pal}\ead[label=e1]{suvra.pal@uta.edu}}
\and
\author[B]{\fnms{Wisdom}~\snm{Aselisewine}\ead[label=e2]{wxa5616@mavs.uta.edu}}
\address[A]{Department of Mathematics, University of Texas at Arlington\printead[presep={,\ }]{e1}}
\address[B]{Department of Mathematics, University of Texas at Arlington\printead[presep={,\ }]{e2}}
\end{aug}

\begin{abstract}

The promotion time cure rate model (PCM) is an extensively studied model for the analysis of time-to-event data in the presence of a cured subgroup. There are several strategies proposed in the literature to model the latency part of PCM. However, there aren't many strategies proposed to investigate the effects of covariates on the incidence part of PCM. In this regard, most existing studies assume the boundary separating the cured and non-cured subjects with respect to the covariates to be linear. As such, they can only capture simple effects of the covariates on the cured/non-cured probability. In this manuscript, we propose a new promotion time cure model that uses the support vector machine (SVM) to model the incidence part. The proposed model inherits the features of the SVM and provides flexibility in capturing non-linearity in the data. To the best of our knowledge, this is the first work that integrates the SVM with PCM model. For the estimation of model parameters, we develop an expectation maximization algorithm where we make use of the sequential minimal optimization technique together with the Platt scaling method to obtain the posterior probabilities of cured/uncured. A detailed simulation study shows that the proposed model outperforms the existing logistic regression-based PCM model as well as the spline regression-based PCM model, which is also known to capture non linearity in the data. This is true in terms of bias and mean square error of different quantities of interest, and also in terms of predictive and classification accuracies of cure. Finally, we illustrate the applicability and superiority of our model using the data from a study on leukemia patients who went through bone marrow transplantation. 

\end{abstract}


\end{frontmatter}

\section{Introduction}\label{sect1}   

Traditional methods of survival analysis, including the Cox's proportional model and the accelerated failure time model, are based on the implicit assumption that all subjects will eventually experience the event of interest (death, relapse, etc.) if the follow-up is sufficiently long. However, if we consider clinical trials with good prognosis, we see that a significant number of subjects respond favourably to the treatment and survive the end of the study without experiencing the event. This leads to heavy censoring towards the end of the study, where the study period is sufficiently long. In such a case, the Kaplan-Meier survival curve shows a long plateau, indicating the presence of cured subjects. To accommodate these cured subjects, along with those subjects who are susceptible to the event, researchers proposed the mixture cure rate model (Boag, 1949; Berkson and Gage, 1952). Since then, the mixture cure rate model has become the most commonly used cure rate model in the literature. Let $Y_1$ denote the time-to-event for a susceptible subject and $Y_0$ denote the time-to-event for a cured subject, where $Y_0$ is such that $P[Y_0=\infty]=1$. Furthermore, let $Y$ denote the time-to-event for any subject in the mixture population (consisting of cured and susceptible subjects). Then, we have $Y=JY_1 + (1-J)Y_0$, where $J$ is an indicator variable that takes the value 1 if a subject is susceptible and takes the value 0 if a subject is cured. The mixture cure rate model can be expressed as:
\begin{equation}
    S_p(y;\bm x,\bm z)=P[Y>y]=(1-\pi(\bm x))+\pi(\bm x) S(y;\bm z),
    \label{mix}
\end{equation}
where $S(y;\bm z)=P[Y_1>y]$ and $\pi(\bm x)=P[J=1]$ with $\bm x$ and $\bm z$ denoting the vectors of covariates affecting the incidence and latency parts of the model, respectively. For a book-length account on mixture cure rate model, interested readers can look at published research monographs (Maller and Zhou, 1996; Peng and Yu, 2021). 

A major drawback of the mixture cure rate model is that it does not have the proportional hazards structure in general. \textcolor{black}{Another drawback of the mixture cure rate model is that it cannot accommodate a practical scenario where an unobserved (or latent) number of cancerous cells, left active after an initial treatment, compete to produce the event of interest, and it is not known which cancerous cell actually caused the event; see Yakovlev and Tsodikov (1996) and Chen et al. (1999). Note that this scenario is different from the traditional competing risks scenario where the number of competing risks and which risk caused the event are both known.} To preserve the proportional hazards structure and to incorporate the aforementioned practical scenario, researchers proposed another approach to define cure rate model that is motivated by the theory of tumor kinetics in cancer studies (Yakovlev and Tsodikov, 1996). \textcolor{black}{Let $N$ denote the latent number of cancerous cells that are related to the occurrence of an event (e.g., recurrence of a cancerous tumor).} For a cured subject, we have $N=0$, whereas for a susceptible subject, we have $N>0$. \textcolor{black}{The susceptible subject will experience the event if one of the $N>0$ cancerous cells get activated and produces the event, where it is not known which cancerous cell actually produced the event.} Let $W_j, j=1,2,\cdots,N,$ denote the promotion time, i.e., the time taken by the $j$-th cancerous cell to produce the event. Then, the time-to-event (or lifetime) under such a scenario is defined as:
\begin{equation}
    Y = \min\{ W_0,W_1,\cdots,W_N\},
    \label{Y}
\end{equation}
where $W_0$ is such that $P[W_0=\infty]=1$. Note that eqn. \eqref{Y} is known as the first activation scheme and it is one of the many activation schemes developed that are suitable for tumor kinetics (Cooner et al., 2007). Conditional on $N$, it is assumed that $W_j$'s, for $j=1,2,\cdots,N$, are independent and they have a common cumulative distribution function (cdf) $F(\cdot|\bm z)$, where $\bm z$ is a set of covariates affecting the promotion times, i.e., the latency part; \textcolor{black}{see Chen et al. (1999) for the model assumptions}. The introduction of covariates to model the cdf of the promotion times readily allows us to capture the heterogeneity in the population under study. If $N$ follows a Poisson distribution with mean $\theta(\bm x)$, the survival function of $Y$, also called the long-term survival function or the population survival function, is given by
\begin{eqnarray}
    S_p(y;\bm x,\bm z) = P[Y>y] &=& P[N=0] + P[W_1>y,W_2>y,\cdots,W_N>y,N>0] \nonumber \\
    &=& \exp(-\theta(\bm x)F(y;\bm z)),
    \label{YP}
\end{eqnarray}
where $\bm x$ is a set of covariates that affect the incidence part, i.e., the mean number of competing risks or, equivalently, the probability of cure. It is important to note that $\bm x$ and $\bm z$ can share common elements. The model in eqn. \eqref{YP} is known in the literature as the promotion time cure rate model (PCM) and has been extensively studied. To estimate the parameters of the PCM model, a Bayesian approach (Chen et al., 1999) and an expectation maximization (EM) algorithm (Chen and Ibrahim, 2001) have been developed. The PCM has also been extended to incorporate longitudinal covariates (Brown and Ibrahim, 2003). While most of the studies on PCM have focused on right censored data, work also exists on the development of the inferential procedures based on interval censored data (Liu and Shen, 2009). From eqn.\eqref{YP}, the cure rate is given by $1-\pi(\bm x) = \lim_{y\rightarrow\infty}S_p(y;\bm x,\bm z) = \exp(-\theta(\bm x))$.

The standard approach to model the effect of covariates on the incidence is to use a log-linear link function for the mean competing risks, i.e., $ \theta(\bm x) = \exp(\bm \gamma^\prime \bm x)$, where $\bm \gamma$ is the vector of regression coefficients corresponding to $\bm x$. Note that $\bm \gamma$ includes the intercept term, which implies that the first element of $\bm x$ is 1. From the above choice of the link function, the covariates are related to the cure rate as: $1-\pi(\bm x) = \exp(-\exp(\bm \gamma^\prime \bm x))$. Equivalently, another approach is to directly link the cure/non-cure rate with the covariates using a sigmoid or logistic function (Balakrishnan and Pal, 2013; Balakrishnan and Pal, 2016) as follows:
\begin{equation}
    \pi(\bm x) = \frac{\exp(\bm \gamma^\prime \bm x)}{1+\exp(\bm \gamma^\prime \bm x)},
    \label{pi}
\end{equation}
which readily implies that $\theta(\bm x) = -\log(1-\pi(\bm x)) = \log(1+\exp(\bm \gamma^\prime \bm x))$. We call this model as the PCM-Logit model. It is important to note that eqn.\eqref{pi} implicitly assumes the boundary separating the cured and non-cured subjects to be linear. Since in a cure rate modeling setup, the information on the cured status is unknown for all subjects whose lifetimes are right censored, the assumption of a linear classification boundary cannot be verified. As such, the use of a logistic link function, as in eqn.\eqref{pi}, is questionable and it may result in biased or imprecise inference for the incidence part, specifically if the true classification boundary is non-linear. Furthermore, such impreciseness may also affect the inference related to the latency part and the overall survival probability. \textcolor{black}{For these reasons, we propose a novel modeling for $\pi(\bm x)$ that employs the support vector machine (SVM) to capture complex effects of covariates $\bm x$ on $\pi(\bm x)$ (i.e., on the incidence).} We call our proposed model as the PCM-SVM model. The integration of SVM and cure rate models is relatively new and is yet to be explored. In this regard, to the best of our knowledge, there is only one recently published article that has used the SVM to model the incidence part of cure rate model (Li et al., 2020). However, the entire formulation is based on the mixture cure rate model. Motivated by this work and given the drawbacks of the mixture cure rate model, we extend the development of Li et al. (2020) to the promotion time cure rate model. For this purpose, we develop an estimation method based on the EM algorithm, where we make use of the sequential minimal optimization (SMO) technique (Platt 1999) and the Platt scaling method (Platt et al., 1999). \textcolor{black}{Furthermore, we compare our proposed PCM-SVM model with the PCM-Logit model as well as the spline-based PCM model (PCM-Spline; see Chen and Du, 2018) and Neural network-based PCM model (PCM-NN; see Xie and Yu, 2021), which are also known to capture complex patterns in the data. We show that when the true classification boundary is non-linear or complex, the proposed PCM-SVM model results in more precise (i.e., smaller bias) and more efficient (i.e., smaller mean square error) estimates of the cured/uncured probabilities when compared to PCM-Logit, PCM-Spline, and PCM-NN models}. Furthermore, we also show that our proposed model results in the highest predictive and classification accuracies. In addition, we show that the ability of our model to capture complex classification boundaries also improves the estimation results related to the latency part and overall population survival. 

The rest of this paper is organized as follows. In Section \ref{sect2}, we discuss the formulation of the proposed PCM-SVM model. In Section \ref{sect3}, we discuss the development of the estimation technique based on the EM algorithm. In Section \ref{sect4}, we present the results of a detailed simulation study to demonstrate the performance and superiority of the proposed model under different scenarios. In Section \ref{sect5}, we illustrate our proposed model using the data from a study on leukemia patients who went through bone marrow transplantation. Finally, in Section \ref{sect6}, we make some concluding remarks and point out some future research directions that integrate machine learning with cure rate models.

\section{Promotion time cure rate model with support vector machine}\label{sect2}

\subsection{Censoring scheme and latency modeling} \label{sect2.1}

    We consider the form of the observed data to be non-informative right censored. Let $Y$ denote the true lifetime and $C$ denote the right censored time. Then, the observed lifetime, denoted by $T$, is given by $T=\min\{Y,C\}$. Let $\delta$ denote the right censoring indicator, defined as: $\delta=1$ if $T=Y$ and $\delta = 0$ if $T=C$. Considering $n$ subjects in the study, the observed data is defined as: $D_0=\{(t_i,\delta_i,\bm x_i,\bm z_i), i=1,2,\cdots,n\}$. Noting that the number of cancerous cells $N_i$, for the $i$-th subject, is unobserved and hence missing, we can define the complete data as: $D_C=\{(t_i,\delta_i,N_i,\bm x_i,\bm z_i), i=1,2,\cdots,n\}$. Next, we focus on modeling the promotion times $W_{ij}, j=1,2,\cdots,N_i$, for the $i$-th subject ($i=1,2,\cdots,n$). For this purpose, we assume a common proportional hazards structure for the promotion times corresponding to the $i$-th subject. Thus, we have the hazard function of the promotion times as:
\begin{equation}
    h(t_i;\bm z_i) = h_0(t_i) \exp(\bm z_i^\prime \bm\beta), 
    \label{haz}
\end{equation}
where $h_0(t_i)$, $i=1,2,\ldots,n$, is the baseline hazard function which can be any arbitrary unspecified hazard function that is not a function of $\bm z_i$, and $\bm\beta$ is the vector of regression coefficients (without the intercept term) corresponding to $\bm z_i$. From eqn.\eqref{haz}, the cdf of the promotion times is given by $F(t_i;\bm z_i)=1-S_0(t_i)^{\exp(\bm z_i^\prime \bm\beta)}$, where $S_0(t_i)$ is the baseline survival function given by $S_0(t_i)=\exp\{-\int_0^{t_i} h_0(u)du \}$. Hence, using the definition of the observed data and the proposed modeling for the promotion times, the promotion time cure rate model, in terms of the population survival function, can be expressed as:
\begin{equation}
    S_p(t_i;\bm x_i,\bm z_i) = \exp\{-\theta(\bm x_i)(1-S_0(t_i)^{\exp(\bm z_i^\prime \bm\beta)}) \},
\end{equation}
\textcolor{black}{where $\theta(\bm x_i) = -\log(1-\pi(\bm x_i))$}. Note that one can also use other modeling approaches for the promotion times such as the accelerated failure time model (Cai et al., 2012) or the piecewise exponential model (Yin and Ibrahim, 2005; Balakrishnan et al., 2016) or a completely parametric model (Balakrishnan and Pal, 2012,2013,2015a,2015b).

\subsection{Modeling the incidence part with SVM}\label{sect2.2}

\textcolor{black}{In this section, we propose novel modeling for $\pi(\bm x)$ using the SVM with the goal of capturing non-linearity in the data. Note that once $\pi(\bm x)$ is obtained, the Poisson mean $\theta(\bm x)$ can be calculated using the relation $\theta(\bm x) = -\log(1-\pi(\bm x))$.} SVM is a type of supervised machine learning algorithm that looks for a hyperplane in a multi-dimensional feature space so as to maximize the margin or the separating space between the two classes (Cortes and Vapnik, 1995; Fouodo et al., 2018). In the given context, the two classes are the cured and the non-cured subjects. The SVM is attractive in the sense that it can separate complex inseparable data by transforming it to a higher dimensional space using some kernel trick. As a result, SVM is considered to be a robust and flexible classifier when compared to the traditional logistic link function or even the probit or the complementary log-log link functions. If we are given a set of covariates $\bm x_i$ (corresponding to the $i$-th subject) that affects the cured or the non-cured probability, and if we know the cured statuses of all subjects, then, the SVM aims to build an optimal decision or classification rule given by
\begin{equation}
   g(\bm x)=\sum_{i=1}^n c_i V_i K_1(\bm x_i, \bm x) - b,
   \label{svm}
\end{equation}
where $V_i=1$ if the $i$-th subject is non-cured (or susceptible) and $V_i=-1$ if the $i$-th subject is cured, $K_1(\cdot,\cdot)$ is a suitable kernel function, $c_i$'s and $b$ are unknown parameters. The kernel function $K_1(\cdot,\cdot)$ is usually chosen to be a symmetric positive semi-definite function. In our case, we will use a radial basis kernel function, which is given by
\begin{equation}
    K_1(\bm x_i, \bm x_j)=\exp\left\{-\frac{(\bm x_i-\bm x_j)^{\prime}(\bm x_i-\bm x_j)}{2\sigma^2}\right\},
\end{equation}
    where the role of the parameter $\sigma$ is to determine the kernel width. One is of course free to use any other kernel functions such as the polynomial or the sigmoid kernel functions. In eqn.\eqref{svm}, the unknown parameters $c_i$, for $i=1,\ldots,n$, are obtained by solving the following optimization problem:
    \begin{equation}
    \underset{c_1, \dots, c_n}{\max}\left[ -\frac{1}{2} \sum_{i=1}^n \sum_{j=1}^n c_i c_j V_i V_j K_1(\bm x_i, \bm x_j) + \sum_{i=1}^n c_i \right].
    \label{opt}
    \end{equation}
    Note that eqn.\eqref{opt} is subject to the constraint $\sum_{i=1}^n c_i V_i=0$ and $0\le c_i \le Q$, for $i=1, \dots, n$, where $Q$ is a tuning parameter that controls model over-fitting. Both parameters $Q$ and $\sigma$ (coming from the kernel function) should be tuned to provide the highest classification accuracy and this can be achieved through cross validation method. For solving eqn.\eqref{opt}, we use the sequential minimal optimization (SMO) method (Platt, 1999), which is a preferred method here as it breaks a large quadratic programming optimization problem into a series of smallest possible quadratic programming problems. The SMO is designed to handle very large training data sets and is time inexpensive. Once we estimate the parameters $c_i$'s, the threshold $b$ can be estimated by
\begin{equation}
b=\sum_{i=1}^n c_i V_i K_1(\bm x_i, \bm x_{j})-V_j,
\end{equation}
for some $c_j > 0$. For a subject having covariate vector $\bm x$, SVM classifies the subject as belonging to the susceptible group if $g(\bm x)>0$ and as belonging to the cured group if $g(\bm x)<0$. Note that to employ the SVM, we have assumed the cured status $V_i$ to be known for all $i=1,2,\cdots,n.$ We know that if $i\in \Delta_1=\{i:\delta_i=1\}$, then, $V_i=1$. However, the values of $V_i$'s are unknown if $i\in \Delta_0=\{i:\delta_i=0\}$. In Section \ref{sect3} below, we will discuss a multiple imputation technique to generate the values of missing $V_i$'s so that the SVM can be easily employed. 

In the context of cure rate model, besides classifying subjects to cured or non-cured groups, it is also of primary interest to estimate the cured or non-cured probabilities. For this purpose, we employ the Platt scaling method (Platt et al., 1999) that uses a method for fitting a sigmoid which maps SVM outputs to posterior probabilities. Using the Platt scaling method, the estimates of the posterior probabilities are given by
\begin{equation}
{\pi(\bm x)} = \frac{1}{1+\exp\{A g(\bm x)+ B\}},
\label{Platt}
\end{equation}
where $A$ and $B$ are the solutions of the following maximization problem:
\begin{equation}
\max_{A,B}\bigg[\sum_{i=1}^n (1- \Tilde{n}_i)[A g(\bm x_i) + B] - \sum_{i=1}^n\log[1 + \exp\{A  g(\bm x_i) + B\}]\bigg].
\label{max}
\end{equation}
In eqn.\eqref{max}, $\Tilde{n}_i$ defined as:
\begin{equation}\label{eq13.8}
  \Tilde{n}_i =
    \begin{cases}
      \frac{n^{(1)}+1}{n^{(1)}+2}, & \text{ if } V_i=1\\
      \frac{1}{n^{(0)}+2}, & \text{ if } V_i=-1,
    \end{cases}       
\end{equation}
where $n^{(1)}$ is the number of non-cured subjects and $n^{(0)}$ is the number of cured subjects with $n=n^{(1)}+n^{(0)}$.

\subsection{\it Tuning the PCM-SVM model}\label{2.3}

\textcolor{black}{We address the issue of over/under fitting in two ways. Firstly, by splitting the data into two sets, namely, the training set and the testing set. The training set is used to train the PCM-SVM model, whereas the testing set is used to test or validate the predictive performance of the final PCM-SVM model. Secondly, we perform a grid-search cross-validation tuning to examine the optimal values of the two most critical hyper-parameters of the PCM-SVM model, namely, the cost parameter (denoted by $Q$ above) and the parameter $\gamma$, where $\gamma = \frac{1}{2\sigma^2}$, when training the PCM-SVM model. The parameter $Q$ is a penalization parameter for mis-classification in the fitted model. A large value of $Q$ implies a substantial penalty for mis-classification and vice versa. The parameter $\gamma$ coming from the radial basis kernel controls the similarity impact of a training point, which is related to the performance of the fitted model. A large value of $\gamma$ implies a small similarity radius and vice versa. With the grid-search cross-validation technique, we specify different plausible values for each hyper-parameter (e.g., $\gamma = \{ 2^{-6},2^{-5},2^{-4}\}$ and $Q = \{ 2^{4},2^5,2^6\}$). The best optimal hyper-parameters obtained through this search are then
used to fit the final PCM-SVM model. 
Finally, we validate the performance of the final fitted PCM-SVM model using the testing data. Model performance evaluation criteria such as the graphical receiver operating characteristic (ROC) curve and it’s area under the curve (AUC) are used to evaluate the performance of the final model.}

\section{Development of the EM algorithm}\label{sect3}

For the promotion time cure rate model with the number of cancerous cells following a Poisson distribution, it can be shown that the joint distribution of the complete data $D_C = \{(t_i,\delta_i,N_i,\bm x_i,\bm z_i), i=1,2,\cdots,n\}$, is given by
\begin{equation}
    f(t_i,\delta_i,N_i;\bm x_i,\bm z_i) = \{1-F(t_i;\bm z_i)\}^{N_i-\delta_i} \{N_i f(t_i;\bm z_i)\}^{\delta_i} p_{N_i}(\bm x_i),  
    \label{joint}
\end{equation}
$t_i>0; \delta_i=0,1; N_i=\delta_i,\delta_i+1,\cdots$, where $f(t_i;\bm z_i)$ is the density function corresponding to $F(t_i;\bm z_i)$ and \textcolor{black}{$p_{N_i}(\bm x_i) = \frac{e^{-\theta(\bm x_i)}\{\theta(\bm x_i)\}^{N_i}}{N_i!}$ with $\theta(\bm x_i) = -\log(1-\pi(\bm x_i))$. Note that for the proposed PCM-SVM model, $\pi(\bm x_i)$ is calculated by first obtaining the SVM classification rule (given by eqn.\eqref{svm}) and then using the Platt scaling method on the classification rule (given by eqn.\eqref{Platt}).}  Using eqn.\eqref{joint} and after some algebra, the complete data log-likelihood function can be expressed as:
\begin{eqnarray}
  l_c &=& \sum_{i=1}^n \log f(t_i,\delta_i,N_i;\bm x_i,\bm z_i) \nonumber \\
  &=& l_{c1} + l_{c2} + K,
\end{eqnarray}
where 
\begin{equation}
    l_{c1} = \sum_{i=1}^n \big[ \log(1-\pi(\bm x_i)) + N_i\log\{ -\log(1-\pi(\bm x_i)) \} \big],
    \label{l1}
\end{equation}

\begin{equation}
    l_{c2} = \sum_{i=1}^n \big [ N_i\log(1-F(t_i;\bm z_i)) + \delta_i \log(h(t_i;\bm z_i))\big]
    \label{l2}
\end{equation}
and 
\begin{equation}
    K = \sum_{i=1}^n \big[ \delta_i\log(N_i) - \log(N_i!)\big ]
\end{equation}
is a constant that is independent of any model parameters. Note that the missing data $N_i, i=1,2,\cdots,n$, are linear in the complete data log-likelihood function. Furthermore, note that $l_{c1}$ involves parameters related to the incidence part only and $l_{c2}$ involves parameters related to the latency part only. This split of the complete log-likelihood function makes the EM algorithm computationally efficient. Furthermore, the proposed approach is different from the traditional approach where the unknown cured status (for the set of censored lifetimes) is considered as the missing data. In such a case, the complete log-likelihood function cannot be separated into incidence and latency parts, and, hence, does not facilitate the integration of the SVM.

In the expectation step or the E-step of the EM algorithm, we take the conditional expectation of the complete data log-likelihood function given the observed data and current values of the parameters. Thus, we need to compute the following:
\begin{equation}
        E[l_c|D_O,\bm\psi^{(s)}] = E[l_{c1}|D_O,\bm\psi^{(s)}] + E[l_{c2}|D_O,\bm\psi^{(s)}] + K^*,
    \label{E-step}
\end{equation}
where $\bm \psi$ denotes the unknown parameter vector, $s$ denotes the $s$-th iteration step of the EM algorithm and $K^*=E[K|D_O,\bm\psi^{(s)}]$ is a constant. Since the $N_i$'s are linear in $l_{c1}$ and $l_{c2}$, the E-step reduces to computing $E[N_i|D_O,\bm\psi^{(s)}]$, for $i=1,2,\cdots,n$. Thus, it is essential to find out the conditional distribution of $N_i$ given the observed data and current values of the parameters. For this purpose, first, we note that the joint distribution of the observed data $D_0=\{(t_i,\delta_i,\bm x_i,\bm z_i), i=1,2,\cdots,n\}$ is given by (after some algebra)
\begin{eqnarray}
    f(t_i,\delta_i;\bm x_i,\bm z_i) &=& \sum_{N_i=\delta_i}^\infty f(t_i,\delta_i,N_i;\bm x_i,\bm z_i)\nonumber \\
    &=& \{-\log(1-\pi(\bm x_i))f(t_i;\bm z_i) \}^{\delta_i}\exp\{ \log(1-\pi(\bm x_i))F(t_i;\bm z_i)\}, \ \ \delta_i = 0,1.
    \label{e1}
\end{eqnarray}
Using eqns.\eqref{joint} and \eqref{e1}, it can be shown after some algebra that the conditional distribution of $(N_i|D_O)$ can be expressed as:
\begin{eqnarray}
  P[N_i=n_i|D_O] &=& \frac{f(t_i,\delta_i,n_i;\bm x_i,\bm z_i)}{f(t_i,\delta_i;\bm x_i,\bm z_i)}\nonumber \\
  &=& \frac{\{-\log(1-\pi(\bm x_i))(1-F(t_i;\bm z_i))\}^{n_i-\delta_i}\exp\{ \log(1-\pi(\bm x_i))(1-F(t_i;\bm z_i))\}}{(n_i-\delta_i)!}, 
  \label{N}
\end{eqnarray}
where $n_i=\delta_i,\delta_i+1,\cdots$ and $\delta_i=0,1$. From eqn.\eqref{N}, it is easy to see that if $\delta_i=0$, the conditional distribution of $N_i$ is Poisson with mean $-\log(1-\pi(\bm x_i))(1-F(t_i;\bm z_i))$. On the other hand, if $\delta_i=1$, the conditional distribution of $N_i$ is shifted Poisson (also called length-biased Poisson) with the length being shifted up by 1. From eqn.\eqref{N}, we have
\begin{eqnarray}
    E[N_i|D_O,\bm\psi^{(s)}] &=& \delta_i - \log(1-\pi(\bm x_i))(1-F(t_i;\bm z_i))|_{\bm\psi=\bm\psi^{(s)}} \nonumber \\
    &=& N_i^{(s)}.
    \label{EN}
\end{eqnarray}
Note that using eqn.\eqref{EN}, we can compute the conditional probability of a subject to be non-cured. Once we obtain $N_i^{(s)}$, we can evaluate the E-step, i.e., eqn.\eqref{E-step}, which reduces to
\begin{equation}
E[l_c|D_O,\bm\psi^{(s)}] = Q_1 + Q_2 + K^*,
\end{equation}
where
\begin{equation}
    Q_1 = \sum_{i=1}^n \big[ \log(1-\pi(\bm x_i)) + N_i^{(s)}\log\{ -\log(1-\pi(\bm x_i)) \} \big]
\end{equation}
and
\begin{equation}
    Q_2 = \sum_{i=1}^n \big [ N_i^{(s)}\log(1-F(t_i;\bm z_i)) + \delta_i \log(h(t_i;\bm z_i))\big]
\end{equation}
with $h(t_i,\bm z_i)$ defined as in eqn.\eqref{haz} and $F(t_i,\bm z_i)=1-S_0(t_i)^{\exp(\bm z_i^\prime \bm\beta)}$.

In the maximization step or the M-step of the EM algorithm, the standard approach in the context of cure model is to maximize $Q_1$ and $Q_2$ independently to obtain updated estimates of $\pi(\bm x_i)$ and $(S_0(\cdot),\bm\beta)$, respectively. However, in this work, we do not maximize the function $Q_1$ to estimate $\pi(\bm x_i)$. Instead, as discussed in Section \ref{sect2.2}, we employ the SVM to first obtain the classification rule $g(\bm x_i)$ (as in eqn.\eqref{svm}) and then use the Platt scaling method (as in eqn.\eqref{Platt}) to estimate the posterior non-cured probabilities $\pi(\bm x_i)$. Now, to employ the SVM, note that we need the values of $V_i, i=1,2,\cdots,n.$ However, as discussed before, the values of $V_i$'s are unobserved for all subjects whose lifetimes are right censored. To solve this issue, we propose to impute the values of $V_i$'s by using a multiple imputation-based technique, \textcolor{black}{along the lines of Li et al. (2020).} For this purpose, at the $s$-th iteration step of the EM algorithm, we discuss two different methods to generate $V_i$'s. For both methods, we need to select a positive integer, say $M$. In method 1, we first generate $\{ N_i^{(k)}, i=1,2,\cdots,n;k=1,2,\cdots,M \}$ by using eqn.\eqref{N}. If the generated $N_i^{(k)}>0$, we set $V_i^{(k)}=1$, else, we set $V_i^{(k)}=-1$. In method 2, we first compute the conditional probability of non-cured by using eqn.\eqref{N}, which is given by
\begin{eqnarray}
    P[V_i = 1|D_O] &=& \delta_i + (1-\delta_i)\{1-P[N_i=0|D_O]\}|_{\bm\psi=\bm\psi^{(s)}} \nonumber \\
    &=& \delta_i + (1-\delta_i)\{1-(1-\pi(x_i))^{1-F(t_i;\bm z_i)}\}|_{\bm\psi=\bm\psi^{(s)}}\nonumber \\
    &=& w_i^{(s)}.
    \label{w}
\end{eqnarray}
Then, we generate $\{ J_i^{(k)}, i=1,2,\cdots,n;k=1,2,\cdots,M \}$ from a Bernoulli distribution with probability of success $w_i^{(s)}$. If $J_i^{(k)}=1$, we set $V_i^{(k)}=1$, else, we set $V_i^{(k)}=-1$. \textcolor{black}{Technically speaking,
both methods are equivalent but the second method is much more easier to implement, which motivated us to employ method 2 in this paper}. For all practical purposes, the value of $M$ can be chosen as 5 (Li et al., 2020). Once we have the generated $V_i^{(k)}, i=1,2,\cdots,n$, for each $k=1,2,\cdots,M$, we estimate $\pi(\bm x_i)$ by employing the SVM followed by the Platt scaling method. We denote these estimates by $\pi^{(k)}(\bm x_i)$. Finally, for each $i=1,2,\cdots,n$, we obtain the estimate of $\pi(\bm x_i)$ as: $\pi(\bm x_i)=\frac{1}{M}\sum_{k=1}^M\pi^{(k)}(\bm x_i)$.

As far as the maximization of the function $Q_2$ is concerned, note that $Q_2$ is similar to the log-likelihood function corresponding to a Cox's proportional hazard model. The only difference is that in $Q_2$ there is an additional constant term $N_i^{(s)}$. Hence, following the work of Peng and Dear (Peng and Dear, 2000), we propose to approximate $Q_2$ by a partial log-likelihood function given by
\begin{equation}
 \sum_{j=1}^{n_k} \log \frac{\exp(\bm s_j^{\prime}\bm \beta)}{\left[\sum_{i' \in R_j}N_{i'}^{(s)}\exp(\bm z_{i'}^{\prime}\bm \beta)\right]^{d_j}},
\label{PD}
\end{equation}
where $\tau_1<\tau_2< \dots< \tau_{n_k}$ are $n_k$ distinct ordered uncensored failure times, $d_j$ denotes number of uncensored failure times equal to $\tau_j$, $R_j$ denotes the risk set at $\tau_j$, and $\bm s_j=\sum_{\{i: t_i=\tau_j\}}\bm z_i$, for $1\le j \le n_k$. Since eqn.\eqref{PD} is independent of the baseline hazard function, we can use the ``coxph()" function in R software to estimate $\bm \beta$ with $\{\log N_i^{(s)}\}$ treated as an offset variable. Note that in the E-step we need to update $N_i^{(s)}$ using eqn.\eqref{EN}. This updating involves $F(t_i;\bm z_i)$, which in turn involves estimation of $S_0(t_i)$ for given estimate of $\bm \beta$. For this purpose, at the $s$-th step, we propose to estimate $S_0^{(s)}(t_i)$ by a Breslow-type estimator given by 
\begin{equation}\label{eq13}
 \hat{S_0}^{(s)}(t_i)=\exp\left(-\sum_{j:\tau_j < t_i}\frac{d_j}{\sum_{i' \in R_j}N_{i'}^{(s)}\exp(\bm z_{i'}^{\prime}\bm \beta^{(s)}) }\right).
\end{equation}
The E-step and the M-step are then continued until some convergence criterion is achieved. In this work, the following convergence criterion is used:
\begin{equation}
    || \boldsymbol\psi^{(s)} - \boldsymbol\psi^{(s-1)} ||^2_2 < \epsilon,
\end{equation}
where $\epsilon$ is some desired tolerance (such as $10^{-3}$ or $10^{-5}$) and $||\cdot||_2$ denotes the $L_2$-norm.

The standard errors of the estimates are not easily available due to the complexity of the proposed EM algorithm. As such, along the lines of Peng and Dear (2000), we propose a bootstrap technique to compute the standard errors. For this purpose, we denote the number of bootstrap samples by $R$, where each bootstrap sample is obtained by sampling with replacement from the original data. Note that the size of each bootstrap sample is the same as the size of the original data. For each bootstrap sample, we use the proposed EM algorithm to estimate the model parameters. This gives us $R$ different estimates for each model parameter. We finally take the standard deviation of the estimates of each model parameter to approximate the standard error. \textcolor{black}{For all practical purposes and along the lines of Li et al. (2020), the number of bootstrap samples $R$ can be chosen as 100.}

To start the iterative EM algorithm, we need to provide initial guess for $\pi(\bm x_i)$  and ($\bm\beta,S_0(t_i)$), $i=1,2,\cdots,n$. For this purpose, first, to come up with an initial guess for $\pi(\bm x_i)$, we consider the right censoring indicator $\delta_i$ as the indicator for the cured status $V_i$. Thus, if $\delta_i=1$, we consider $V_i=1$, whereas if $\delta_i=0$, we consider $V_i=-1$. Then, we apply the SVM followed by the Platt scaling method to obtain $\pi(\bm x_i)$. Next, to come up with an initial guess for the latency parameters $(S_0(t_i),\bm\beta)$, we can set the starting value of each component of $\bm \beta$ at 0.5 and estimate $S_0(t_i)$ by fitting a Cox proportional hazards model without any covariates.

\section{Simulation study}\label{sect4}

\begin{figure}[ht!]
\begin{tabular}{ccc}
		\includegraphics[scale=0.32]{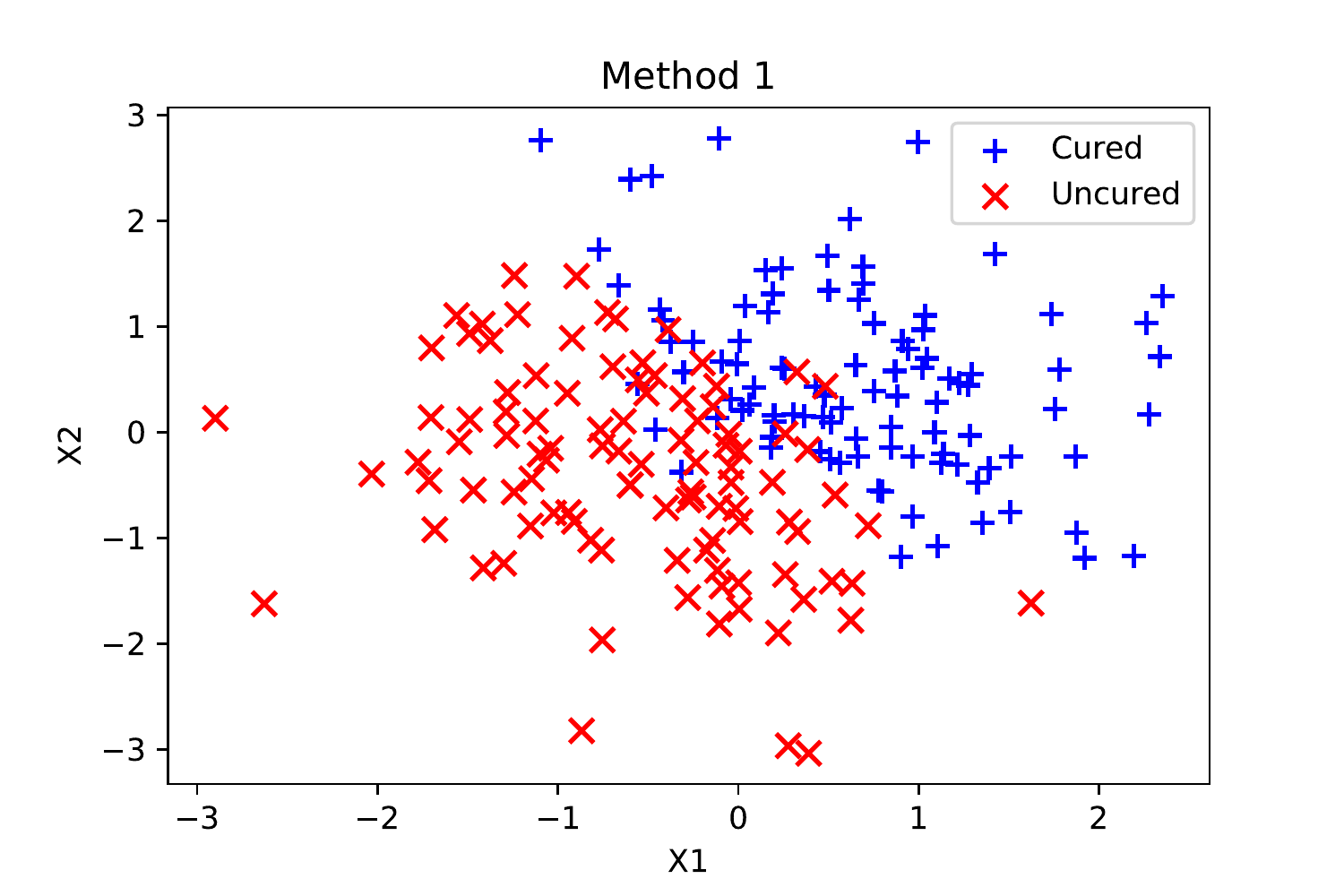}
		\includegraphics[scale=0.32]{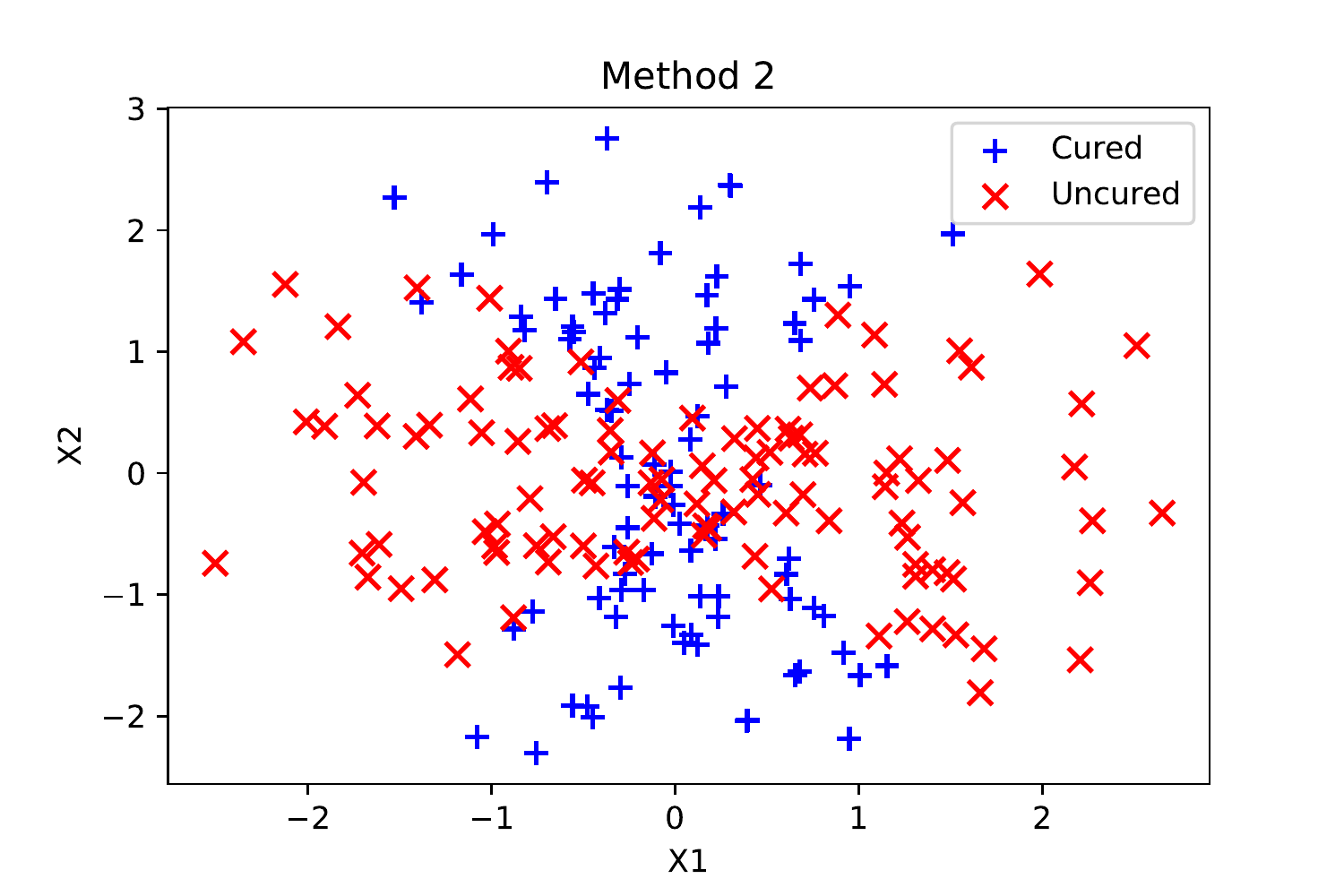}
		\includegraphics[scale=0.32]{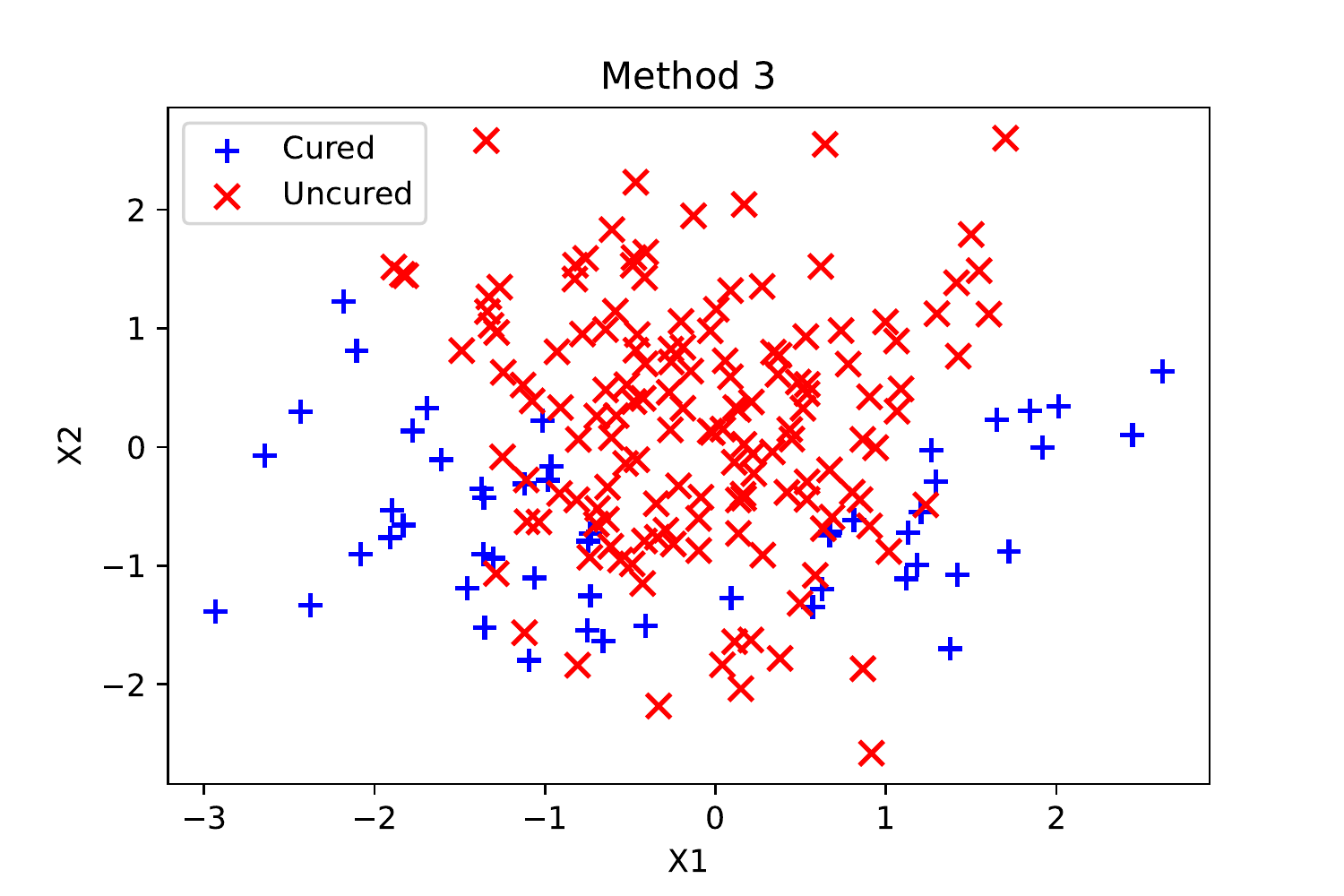}\\
\end{tabular}
  \caption{Simulated cured and non-cured statuses under the three methods}
\label{figure:F1}
\end{figure}

In this section, we demonstrate the performance of the proposed PCM-SVM model and the EM algorithm through a detailed Monte Carlo simulation study. We also compare the performance of the PCM-SVM model with that of the PCM-Logit model through the calculated bias and the mean square error (MSE) of the estimates of the model parameters. We consider different sample sizes as: $n=300$ and $n=600$. \textcolor{black}{Two-third of the data is used as the training set, whereas the remaining one-third of the data is used as the testing set.} To introduce covariates, we generate $x_1$ and $x_2$ independently from the standard normal distribution and we assume $\bm x=\bm z=(x_1,x_2)$. Then, we consider the following three methods to generate the true non-cured probabilities:
\begin{eqnarray*}
\text{Method 1:} \ \ \pi(\boldsymbol x) &=& \frac{\exp(0.3-5x_1-3x_2)}{1+\exp(0.3-5x_1-3x_2)},\\
\text{Method 2:} \ \ \pi(\boldsymbol x) &=&  \frac{\exp(0.3+5x_1^2-3x_2^2)}{1+\exp(0.3+5x_1^2-3x_2^2)},\ \ \text{and} \\
\text{Method 3:} \ \ \pi(\boldsymbol x) &=& \exp\{-\exp(0.3-5\cos x_1-3\sin x_2)\}.
\end{eqnarray*} 
It is easy to see that Method 1 represents the standard logistic link function, which implies that the cured and the non-cured subjects are linearly separable. On the other hand, Methods 2 and 3 are non-logistic functions and can capture more complex classification boundaries. Figure \ref{figure:F1} presents the plots of different simulated classification boundaries. 


For the latency part, we assume the promotion time for each risk factor to follow a proportional hazards model given by (note that $x_1=z_1$ and $x_2=z_2$)
\begin{equation*}
    h(t;\bm z) = h_0(t)\exp(\beta_1 z_1 + \beta_2 z_2),
\end{equation*}
    where we consider the baseline hazard function to be of the following form: $h_0(t)=\alpha t^{\alpha-1}$, $\alpha>0$. Note that the above hazard function with the chosen form for the baseline hazard function implies that the promotion times are distributed as Weibull with shape parameter $\alpha$ and scale parameter $\{\exp(\beta_1 z_1 + \beta_2 z_2) \}^{-1/\alpha}$. We consider the true values of $(\alpha,\beta_1,\beta_2)$ as $(2,1,0.5)$. The censoring times are generated from an exponential distribution with rate parameter 0.2. \textcolor{black}{With these choices of parameters, the censoring proportions for methods 1, 2 and 3 are roughly 0.60, 0.70 and 0.30, respectively, whereas the corresponding cured proportions are respectively around 0.50, 0.60 and 0.20. Thus, methods 1, 2 and 3 correspond to mid, high, and low censoring cases.} Next, to generate the lifetime data $\{(t_i,\delta_i), i=1,\ldots,n$\}, we perform the following steps for the $i$-th subject in the study:
\begin{itemize}
    \item [] Step 1: Generate $U_i$ from a Uniform (0,1) distribution and censoring time $C_i$ from an exponential distribution with rate parameter 0.2.
    \item [] Step 2: If $U_i\leq 1-\pi(\bm x_i)$, set $t_i=C_i$ and $\delta_i=0$.
    \item [] Step 3: If $U_i > 1-\pi(\bm x_i)$, perform the following:
    \begin{itemize}
        \item [3a:] Generate $U_{1i}$ from Uniform$(1-\pi(\bm x_i),1)$ and equate $S_p(y_i)=U_{1i}$, i.e., equate $\{1-\pi(\bm x_i) \}^{F(y_i;\bm z_i)}=U_{1i}$.
        \item [3b:] Solve for $y_i$ as: $y_i = F^{-1}\big\{\frac{\log U_{1i}}{\log(1-\pi(\bm x_i))} \big\}$, where $F(\cdot)$ follows a Weibull distribution with shape parameter $\alpha$ and scale parameter $\{\exp(\beta_1 z_{1i}+\beta_2 z_{2i})\}^{-1/\alpha}$.
        \item [3c:] Set $t_i=\min\{ y_i,C_i\}$.
        \item [3d:] If $t_i=y_i$, set $\delta_i=1$, else, set $\delta_i=0$.
    \end{itemize}
\end{itemize}
All simulations are conducted using the R statistical software (version 4.0.5) and all simulation results are based on $N_1 = 500$ Monte Carlo runs. To employ the SVM, we consider the number of imputations to be 5 to carry out the multiple imputation-based technique to estimate $\pi(\bm x_i)$. Note that this is along the lines of Li et al. (2020). The R codes for the data generation and the SVM-based EM algorithm are provided in the Supplementary Material (Pal and Aselisewine, 2023).

\subsection{Simulation results: incidence part}\label{sect4.2}

In Table \ref{table:T1}, we report the biases and MSEs of the estimates of the non-cured probabilities $\pi(\bm x)$ obtained from the proposed PCM-SVM model and compare them with those obtained from the standard PCM-Logit model. The formulae used to compute the bias and MSE of the estimate of $\pi(\bm x)$, denoted by $\widehat{\pi(\bm x)}$, are respectively given by
\begin{equation}
\text{Bias}(\hat{\pi}(\boldsymbol x)) = \frac{1}{N_1}\sum_{r=1}^{N_1} \bigg [ \frac{1}{n}\sum_{i=1}^n \big\{ \widehat{\pi^{(r)}}(\boldsymbol x_i) - \pi^{(r)}(\boldsymbol x_i) \big\} \bigg]
\label{B1}
\end{equation}
and
\begin{equation}
\text{MSE}(\hat{\pi}(\boldsymbol x)) = \frac{1}{N_1}\sum_{r=1}^{N_1} \bigg [ \frac{1}{n}\sum_{i=1}^n \big\{ \widehat{\pi^{(r)}}(\boldsymbol x_i) - \pi^{(r)}(\boldsymbol x_i) \big\}^2 \bigg],
\label{B2}
\end{equation}
where $\pi^{(r)}(\bm x_i)$ and $\widehat{\pi^{(r)}}(\boldsymbol x_i)$ are the true non-cured probability and the estimated non-cured probability, respectively, corresponding to the $i$-th subject and $r$-th Monte Carlo run, for $i=1,\ldots,n$ and $r=1,\ldots,N_1$. From the results in Table \ref{table:T1}, it is easy to see that when the cured and non-cured subjects are separated by non-linear boundaries (Methods 2 and 3), the biases and MSEs of the estimated non-cured probabilities from the proposed PCM-SVM model is smaller than those from the PCM-Logit model. On the other hand, when the true classification boundary is linear (Method 1), the PCM-Logit model produces smaller MSE in the estimated non-cured probability. This is not surprising since it is expected that a logistic regression-based model will better capture a linear classification boundary. Note, however, that the biases of the estimated non-cured probabilities obtained from both models are similar with the PCM-SVM model producing smaller bias. The findings from Table \ref{table:T1} clearly demonstrates that the proposed PCM-SVM model can capture complex relationships between the non-cured probability and the covariates better than the PCM-Logit model, which in turn improves the accuracy and precision in the estimation of the non-cured probabilities. Note that a larger sample size helps in the reduction of both bias and MSE (with a few exceptions noted under the PCM-Logit model). In the Supplementary Material (Pal and Aselisewine, 2023), we present the biases of the estimates of the non-cured probabilities plotted against both covariates.

\begin{table}[ht!]
\color{black}
\caption{Comparison of bias and MSE of $\widehat{\pi}(\bm x)$}
\centering{
\begin{tabular}{cccc|cc}
\hline
\multirow{3}{*}{$n$} & \multirow{3}{*}{Method} & \multicolumn{4}{c}{$\widehat{\pi}(\bm x)$}   \\ \cline{3-6}
& & \multicolumn{2}{c}{Bias} & \multicolumn{2}{c}{MSE}  \\ \cline{3-6} 
& & PCM-SVM & PCM-Logit & PCM-SVM & PCM-Logit \\ \hline
\multirow{3}{*}{300} & 1 & -0.0021 & -0.0032 & 0.0202 & 0.0024   \\
                     & 2 & -0.0014 & 0.0022 & 0.0286 & 0.1363   \\
                     & 3 & -0.0008 & 0.0023 & 0.0227 & 0.0905   \\ \hline
\multirow{3}{*}{600} 
                     & 1 & -0.0013 & -0.0032 & 0.0140 & 0.0013   \\
                     & 2 & 0.0004 & 0.0033 & 0.0201 & 0.1368     \\
                     & 3 &0.0004 & 0.0032 & 0.0170 & 0.0916 \\ \hline
\end{tabular}
}
\label{table:T1}
\end{table}


\subsection{Simulation results: latency part and survival probabilities}\label{sect4.3}

In Table \ref{table:T2}, we report the biases and MSEs of the estimates of the overall survival probabilities $S_p(\cdot;\bm x,\bm z)$ and susceptible survival probabilities $S_u(\cdot;\bm x,\bm z)$ obtained from the proposed PCM-SVM model and compare them with those obtained from the PCM-Logit model. The susceptible survival probability is defined as $S_u(\cdot;\bm x,\bm z) = \frac{S_p(\cdot; \bm x,\bm z)-(1-\pi(\bm x))}{\pi(\bm x)}$. Similarly, in Table \ref{table:T3}, we report the biases and MSEs of the estimates of the promotion time survival probabilities $S(\cdot;\bm z)$ obtained from both models. Note that $S(\cdot;\bm z)$ depends only on the latency parameters, i.e., $S_0(\cdot)$ and $\bm \beta$. The formulae used to calculate the biases and MSEs of the estimates of the aforementioned survival probabilities are similar to the ones presented in eqns.\eqref{B1} and \eqref{B2} with the uncured probability being replaced by the survival probability of interest. From Tables \ref{table:T2} and \ref{table:T3}, we can see that the performance of the PCM-Logit model is better than the PCM-SVM model (in terms of smaller biases and MSEs in the estimated survival probabilities) only when the true classification boundary is linear, i.e., when Method 1 is true. On the other hand, when Methods 2 and 3 are true, i.e., the true boundary separating the cured and non-cured subjects is non-linear, the biases and MSEs in the estimated survival probabilities obtained from the PCM-SVM model are always smaller. These findings allow us to confirm that the proposed PCM-SVM model's ability to capture non-linearity in the incidence also improves the estimation results for the overall and latency survival probabilities. In the Supplementary Material (Pal and Aselisewine, 2023), we present the biases of the estimates of the susceptible survival probabilities plotted against both covariates. We also present the biases of the estimates of the overall survival probabilities and progression time survival probabilities plotted against the covariates.

\begin{figure}[htb!]
\centering
\begin{tabular}{ccc}
		\includegraphics[scale=0.35]{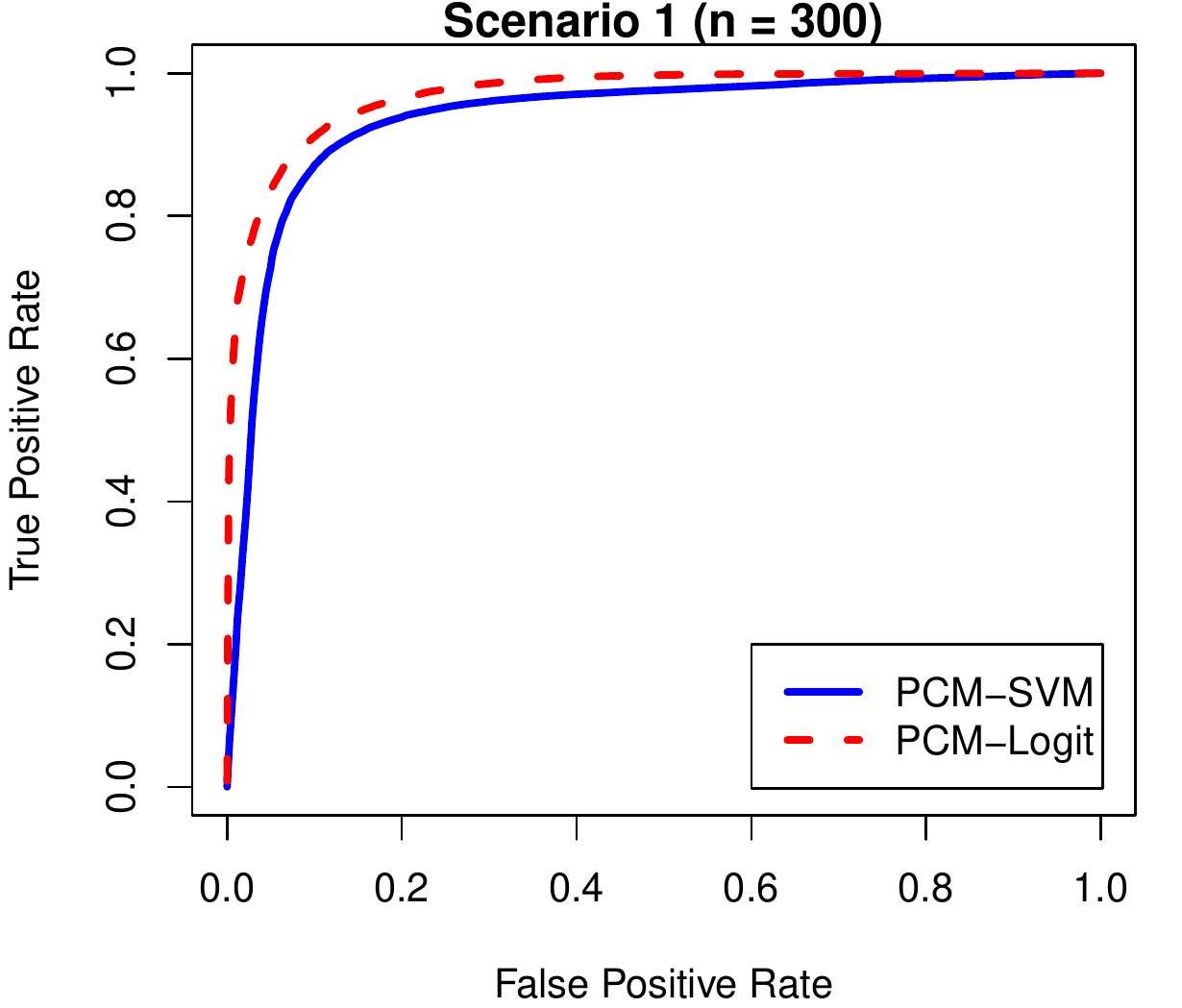}
		\includegraphics[scale=0.35]{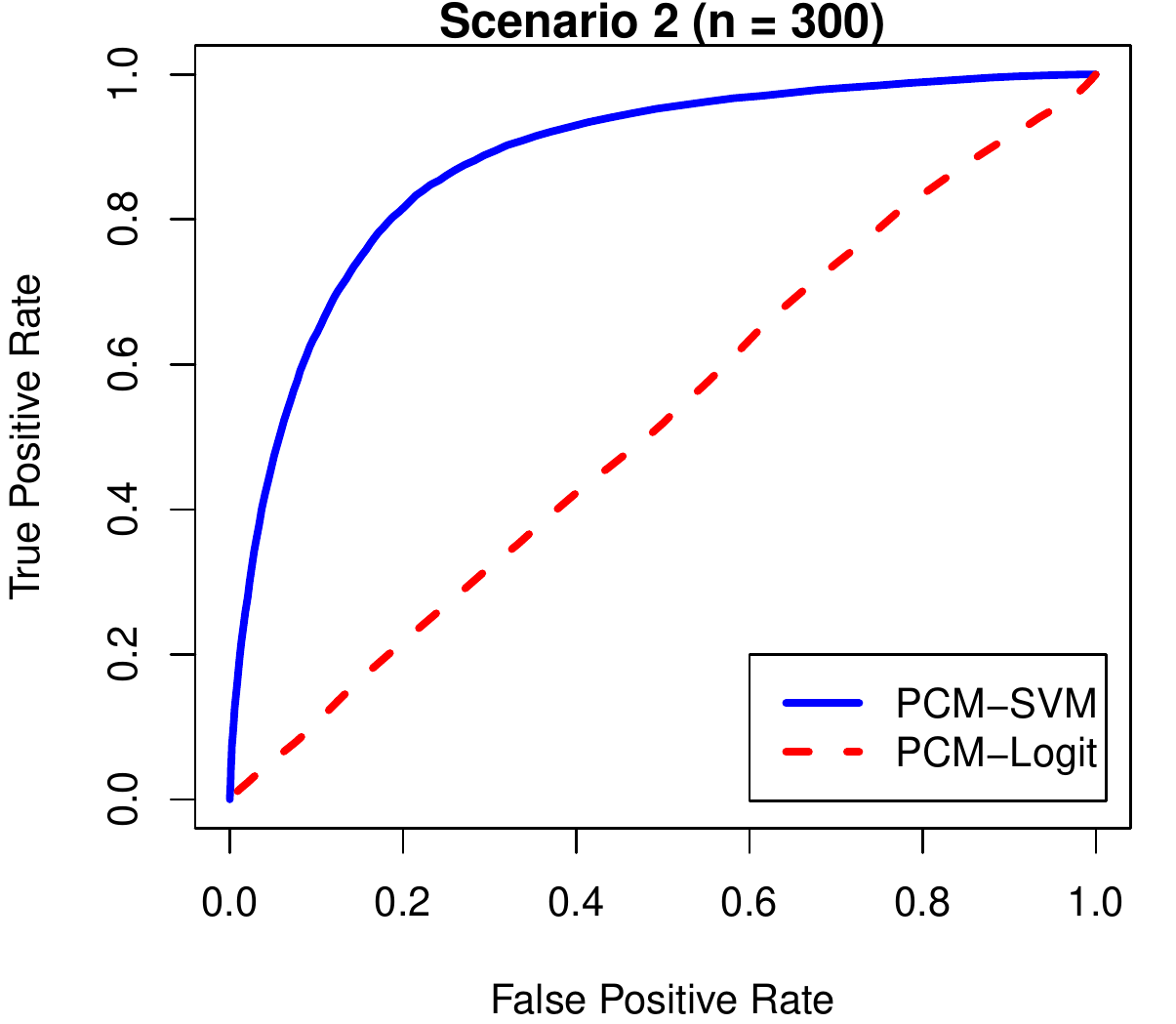}
		\includegraphics[scale=0.35]{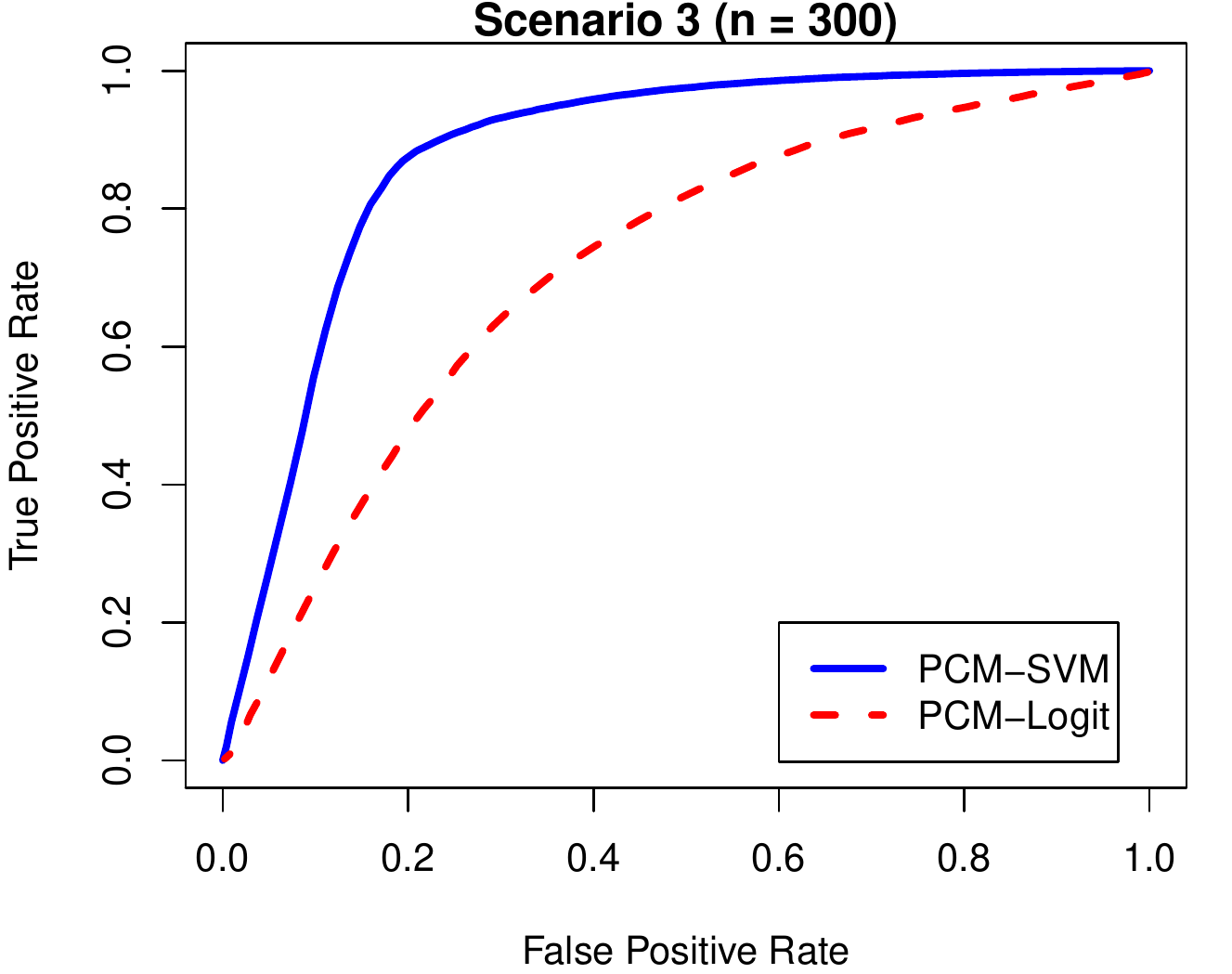}\\
		\includegraphics[scale=0.35]{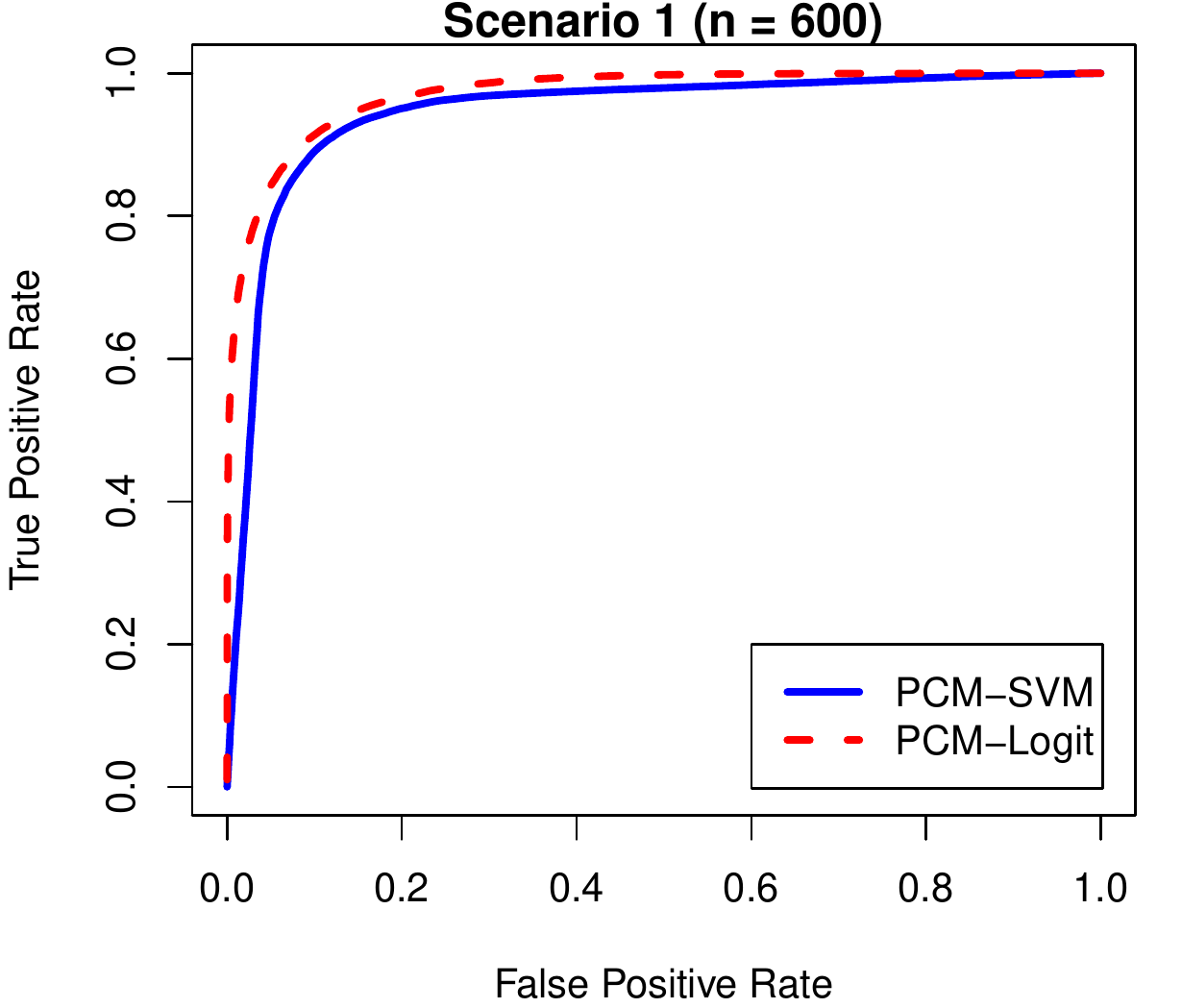}
		\includegraphics[scale=0.35]{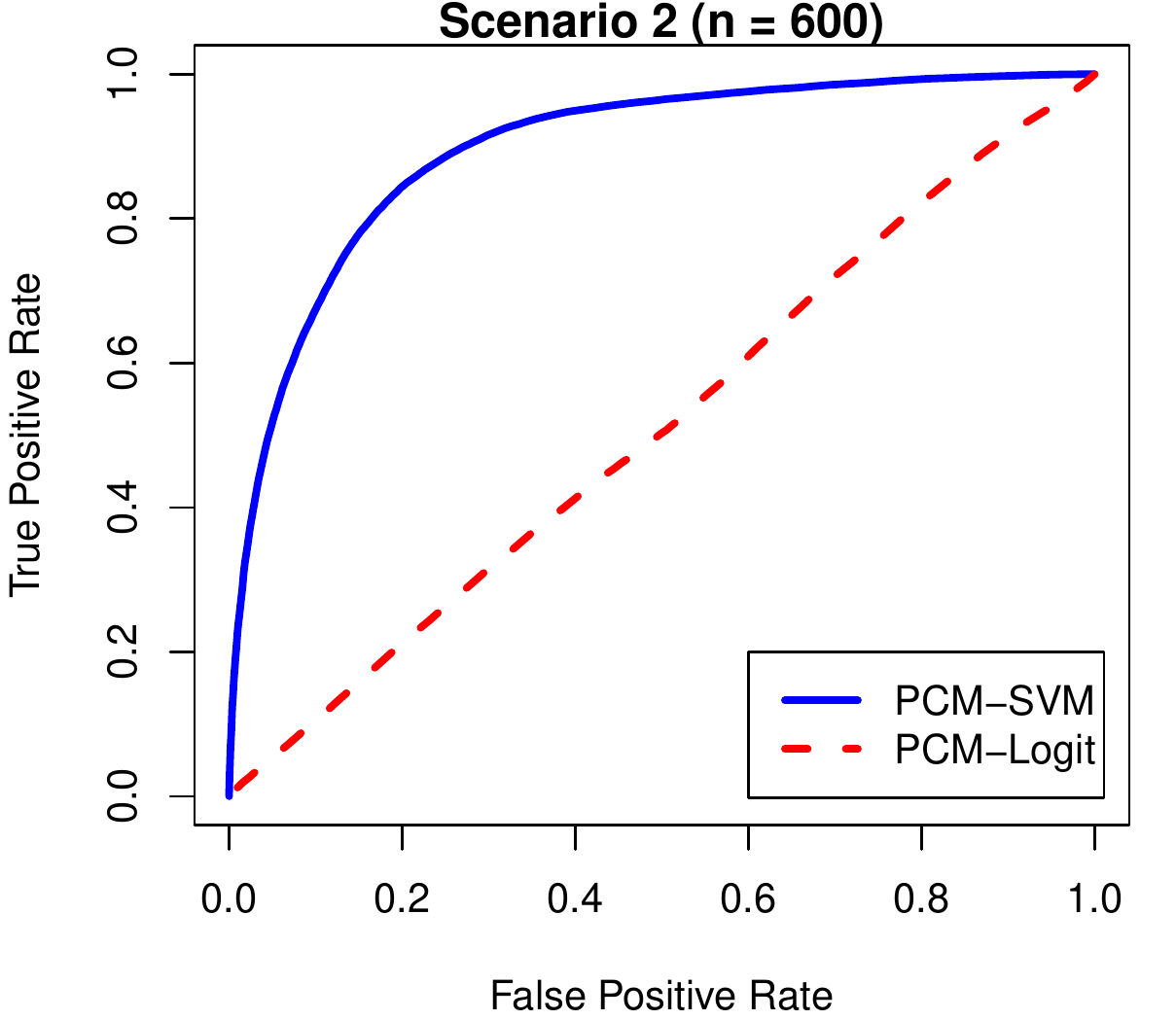}
		\includegraphics[scale=0.35]{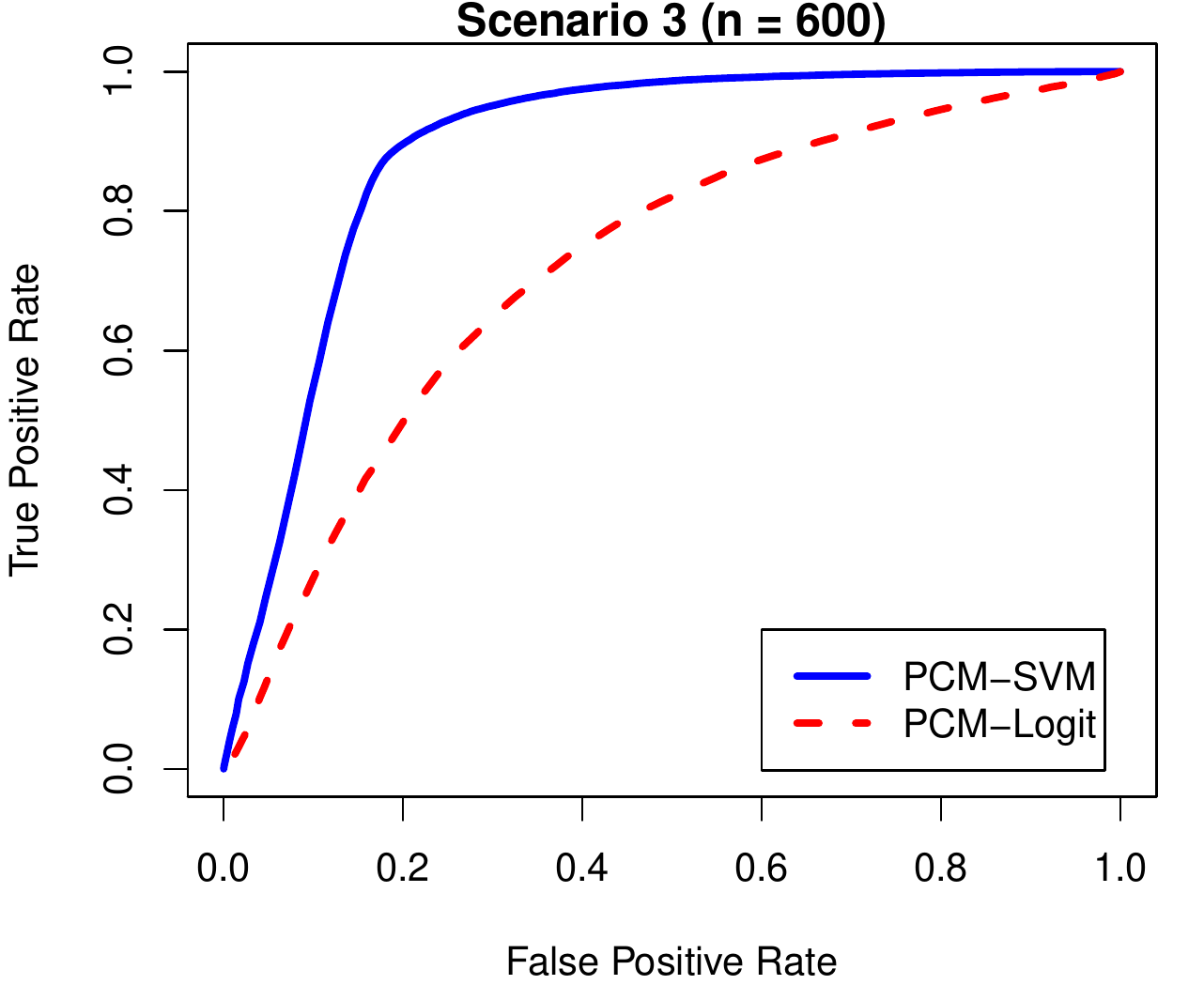}\\
\end{tabular}
  \caption{ROC curves based on a Monte Carlo simulation study for the PCM-SVM and PCM-Logit models.}
\label{figure:ROC-Sim}
\end{figure}

We are aware that in practice the cured statuses are unknown for all subjects whose time-to-events are right censored. However, in a simulation study, it is possible to determine if a right censored observation can be treated as cured. Using such information on the cured statuses, it is of interest to calculate the percentage of correctly classified observations or classification accuracy (CA) for the PCM-SVM model and compare these percentages with those obtained from the PCM-Logit model. For both PCM-SVM and PCM-Logit models, we consider a subject to belong to the susceptible group if its predicted non-cured probability is greater than 0.5. Otherwise, we consider the subject to belong to the cured group. In this way, the comparison between the two competing models is fair. It is also of interest to compare the PCM-SVM and PCM-Logit models using the ROC curves and the corresponding AUC values. In Table \ref{table:T4}, we report the percentage classification accuracies (CA) and the AUC values under different methods that we have considered. Figure \ref{figure:ROC-Sim} presents the ROC curves.  \textcolor{black}{Note that the true label for calculating the AUC and CA is the true cure index for each subject when the data is generated.} From the results in Table \ref{table:T4}, it is once again clear that under complex (non-linear) classification boundaries, PCM-SVM model outperforms the PCM-Logit model with noticeable differences in the AUC and CA values. Under linear classification boundary, the performance of the PCM-SVM model is very close to that of the PCM-Logit model. \textcolor{black}{The closeness of the training and testing AUC values implies that there is no issue with model over-fitting.}

\begin{table}[ht!]
\color{black}
\caption{Comparison of bias and MSE of $\widehat{S_p}(\cdot;\bm x,\bm z)$ and $\hat{S_u}(\cdot;\bm x,\bm z)$}
\resizebox{0.98\textwidth}{!}{
\begin{tabular}{cccc|cc|cc|cc}
\hline
\multirow{3}{*}{$n$} & \multirow{3}{*}{Method} & \multicolumn{4}{c}{$\widehat{S_p}(\cdot;\bm x,\bm z)$} & \multicolumn{4}{c}{$\hat{S_u}(\cdot;\bm x,\bm z)$} \\ \cline{3-10}
& & \multicolumn{2}{c}{Bias} & \multicolumn{2}{c}{MSE} & \multicolumn{2}{c}{Bias} & \multicolumn{2}{c}{MSE} \\ \cline{3-10} 
& & PCM-SVM & PCM-Logit & PCM-SVM & PCM-Logit & PCM-SVM  & PCM-Logit & PCM-SVM & PCM-Logit \\ \hline
\multirow{3}{*}{300} & 1 & -0.0084 & 0.0005 & 0.0131 & 0.0029 & 0.0283 & -0.0007 & 0.0150 & 0.0030 \\
                     & 2 & -0.0057 & -0.0562 & 0.0225 & 0.0810 & -0.0084 & -0.0180 & 0.0111 & 0.0152    \\
                     & 3 & -0.0055 & -0.0360 & 0.0192 & 0.0653 & -0.0004 & -0.0074 & 0.0150 & 0.0147   \\ \hline
\multirow{3}{*}{600} 
                     & 1 & -0.0058 & 0.0014 & 0.0096 & 0.0015 & 0.0302 & 0.0004 & 0.0130 & 0.0015   \\
                     & 2 & -0.0037 &-0.0578 & 0.0171 & 0.0818 &-0.0047 &-0.0170 & 0.0081 & 0.0132 \\
                     & 3 & -0.0024 & -0.0362& 0.0158 & 0.0658 & 0.0046 & -0.0053& 0.0136 & 0.0130  \\ \hline
\end{tabular}
}
\label{table:T2}
\end{table}

\begin{table}[ht!]
\color{black}
\caption{Comparison of bias and MSE of $\widehat{S}(\cdot;\bm z)$}
\centering{
\begin{tabular}{cccc|cc}
\hline
\multirow{3}{*}{$n$} & \multirow{3}{*}{Method} & \multicolumn{4}{c}{$\widehat{S}(\cdot;\bm z)$}   \\ \cline{3-6}
& & \multicolumn{2}{c}{Bias} & \multicolumn{2}{c}{MSE}  \\ \cline{3-6} 
& & PCM-SVM & PCM-Logit & PCM-SVM & PCM-Logit \\ \hline
\multirow{3}{*}{300} & 1 & -0.0074 & -0.0032 & 0.0259 & 0.0028   \\
                     & 2 & -0.0279 & -0.0603 & 0.0074 & 0.0142   \\
                     & 3 & -0.0472 & -0.0826 & 0.0127 & 0.0161   \\ \hline
\multirow{3}{*}{600} 
                     & 1 & -0.0032 & -0.0040 & 0.0221 & 0.0014   \\
                     & 2 & -0.0209 & -0.0597 & 0.0040 & 0.0113     \\
                     & 3 & -0.0407 & -0.0814 & 0.0101 & 0.0142   \\ \hline
\end{tabular}
}
\label{table:T3}
\end{table}

\begin{table}[ht!]
\color{black}
\center
\caption{Comparison of PCM-SVM and PCM-Logit models using AUC and CA (in \%)}
\begin{tabular}{cccc|cc|cc}
\hline
\multirow{2}{*}{$n$} & \multirow{2}{*}{Method} & \multicolumn{2}{c}{Training AUC}& \multicolumn{2}{c}{Testing AUC} & \multicolumn{2}{c}{CA (\%)}\\ \cline{3-8}
                & & PCM-SVM & PCM-Logit   & PCM-SVM & PCM-Logit        & PCM-SVM & PCM-Logit \\ \hline
\multirow{3}{*}{300} & 1 & 0.9683 & 0.9733  &0.9392&0.9722     & 86.00 & 91.00   \\
                     & 2 & 0.9339 & 0.5434  &0.8824&0.5448     & 77.00 & 53.00   \\
                     & 3 & 0.9502 & 0.7376  &0.8845&0.7220     & 88.00 & 73.00   \\ \hline
\multirow{3}{*}{600} 
                     & 1 & 0.9634 & 0.9729  &0.9441&0.9724   & 92.00 & 91.50    \\
                     & 2 & 0.9309 & 0.5300  &0.8969&0.5298   & 88.00 & 57.00     \\
                     & 3 & 0.9357 & 0.7348  &0.8898&0.7241   & 90.00 & 77.00   \\ \hline
\end{tabular}
\label{table:T4}
\end{table}

\color{black}

\subsection{Comparison with spline-based and neural network-based models}\label{sect4.1}

In this section, we compare the performance of the proposed PCM-SVM model with the spline-based (PCM-Spline) and recently proposed neural network-based (PCM-NN) models, which are also capable of capturing non-linearity in the data; see Chen and Du (2018) and Xie and Yu (2021). For this purpose, we use a 10-dimensional covariate space and consider complicated interaction terms along with correlation among covariates. The true non-cured probability is generated using the following:
\begin{eqnarray*}
 \pi(\bm x)&=&1-\exp[-\exp\{0.4(0.05x_1^{2}+0.05\text{tanh}(x_2)\\
 &&-0.05x_3x_4(4-0.0005x_3x_4)(4-0.0005x_3x_4)\\
&& + \log(|x_1 + x_5|))+0.05x_6^{2}+0.05\text{tanh}(x_7)\\
&&-0.05x_8x_9(4-0.0005x_8x_9)(4-0.0005x_8x_9)\\
&&+\log(|x_6+x_{10}|)\}].
\end{eqnarray*}
In the above, $x_1, x_2,\cdots,x_5$ are generated from a multivariate normal distribution, $N(0,\Sigma)$, with $\Sigma = (1, 0.8, 0.5, 0.2, 0; 0.8, 1, 0.2, 0.6, 0; 0.5, 0.2, 1, 0.3, 0; 0.2, 0.6, 0.3, 1, 0; 0, 0, 0, 0, 1)$, whereas $x_6, x_7, \cdots,x_{10}$ are independently generated from the standard normal distribution. Also, we assume $\bm z = \bm x$, i.e., the same set of covariates are assigned to the incidence and latency parts. For the latency, the true values of $(\alpha,\beta_1,\beta_2,\cdots,\beta_{10})$ are chosen as $(3.5,0.8,1.2,0.5,1.1,-0.6,-1.4,-0.5,-0.8,0.5,1.8)$. These parameter settings are along the lines of Xie and Yu (2021). The censoring proportion under this method turns out to be roughly 0.60, whereas the cured proportion is roughly 0.45. From the results presented in Table \ref{table:T5}, it is clear that our proposed PCM-SVM outperforms both PCM-Spline and PCM-NN in terms of bias and MSE when it comes to estimation of the non-cured probability $\widehat{\pi}(\bm x)$ and overall survival probability $\widehat{S_p}(\cdot;\bm x,\bm z)$. When it comes to estimation of the latency survival functions, $\hat{S_u}(\cdot;\bm x,\bm z$) and $\widehat{S}(\cdot;\bm z)$, the performance of the PCM-SVM is considerably better than the PCM-NN (except in one case), whereas the performances of the PCM-SVM and PCM-Spline are comparable. 

\begin{table}[htb!]
\color{black}
\center
\caption{\textcolor{black}{Comparison of PCM-SVM with PCM-Spline and PCM-NN}}
\begin{tabular}{cccc|cc|cc|cc}
\hline
$n$ & Method &\multicolumn{2}{c}{$\widehat{\pi}(\bm x)$}&\multicolumn{2}{c}{$\widehat{S_p}(\cdot;\bm x,\bm z)$} & \multicolumn{2}{c}{$\hat{S_u}(\cdot;\bm x,\bm z)$}&\multicolumn{2}{c}{$\widehat{S}(\cdot;\bm z)$}  \\  \cline{3-10}
& & Bias& MSE    & Bias& MSE   & Bias& MSE &Bias&MSE \\ \hline
\multirow{3}{*}{300}& PCM-SVM   &0.0000  &0.0728  &-0.0244  &0.0472 &-0.0156&0.0217 &-0.0452&0.0199       \\
                    & PCM-Spline&0.0049 &0.0924  &-0.0264  &0.0565 &-0.0142&0.0221 &-0.0368&0.0204\\
                    & PCM-NN    &-0.0274&0.1615  &0.0528   &0.0949 &0.0104&0.0283&0.0744&0.0322\\ \hline
\multirow{3}{*}{600}& PCM-SVM   &-0.0015&0.0681  &-0.0190  &0.0434 &-0.0130&0.0140&-0.0408&0.0110   \\
                    & PCM-Spline&0.0070 &0.0854  &-0.0294  &0.0532 &-0.0118&0.0140&-0.0383&0.0111\\
                    & PCM-NN    &-0.0207&0.1421  &0.0212   &0.0876 &0.0210&0.0255&0.0963&0.0361\\ \hline
\end{tabular}
\label{table:T5}
\end{table}


\color{black}

\section{Application to leukemia data} \label{sect5}

\begin{figure}[ht!]
\centering
\includegraphics[scale=0.8]{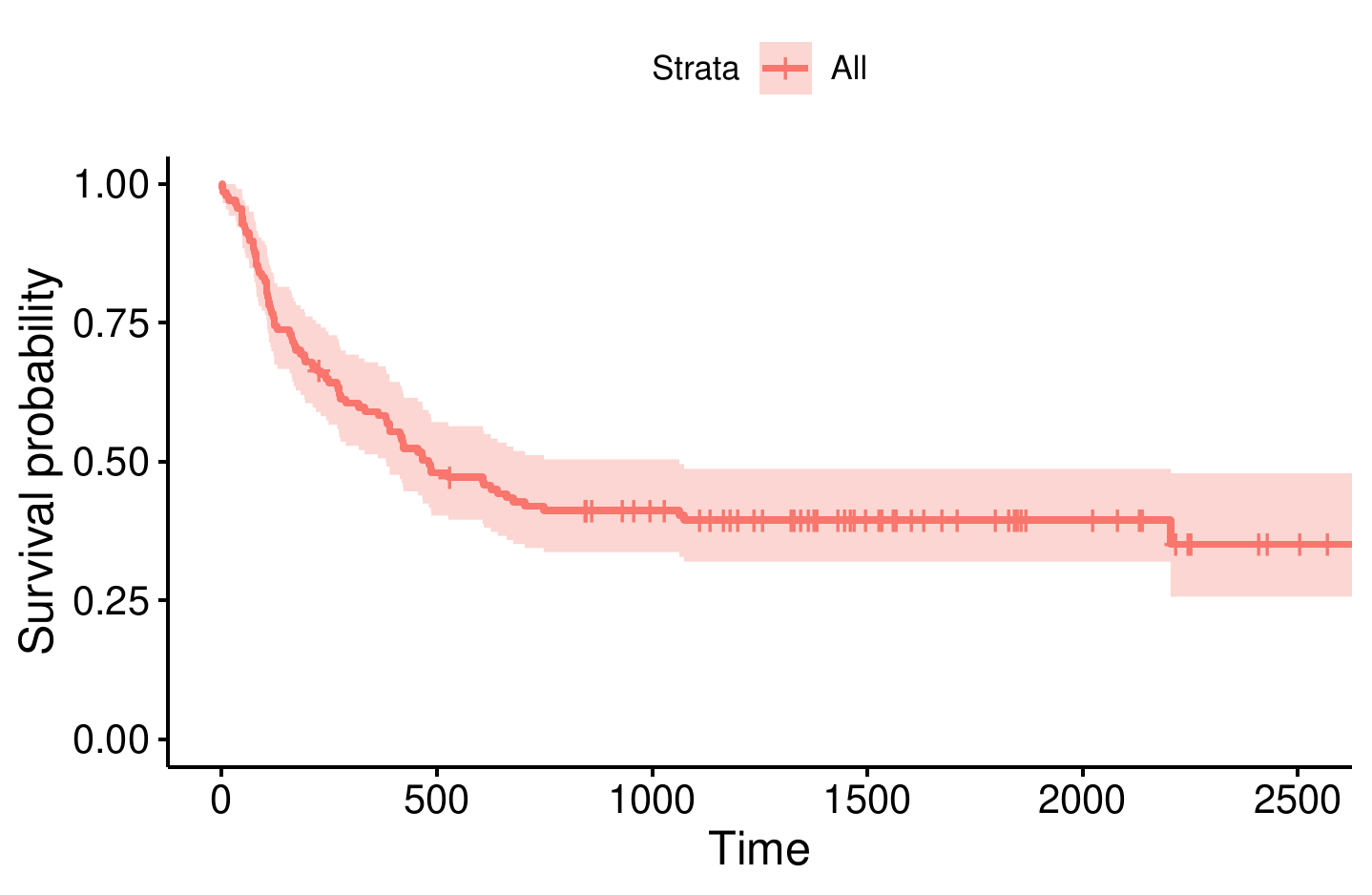}
\caption{Kaplan-Meier survival curve for the leukemia data}
\label{figure:KM}
\end{figure}

In this section, using a real data, we illustrate the performance of the proposed PCM-SVM model and the EM algorithm. For this purpose, we use the data from a study on leukemia patients who went through bone marrow transplantation (Copelan et al., 1991). The complete data set can be downloaded from the R package ``KMsurv". There were 137 leukemia patients registered in the study and these patients were followed up to 2640 days. We consider the event of interest to be relapse or death due to leukemia following bone marrow transplantation. At the end of the study, 54 patients were still alive (disease free survival) and hence were right censored, resulting in 39.4\% censoring. \textcolor{black}{Now, since leukemia is caused by the rapid production of abnormal white blood cells, or cancerous cells, these cancerous cells are obviously related to relapse or death due to leukemia following bone marrow transplantation. Furthermore, we do not observe the number of abnormal white blood cells produced by nature. Thus, the biological process generating the lifetimes of leukemia patients matches the setup considered in this manuscript, and hence the leukemia dataset can be used as an illustrative example.} In this application, we consider the effects of patient's age (in years) and donor's age (in years) on both incidence and latency parts. We first standardize these covariates (by subtracting the mean and dividing by the standard deviation) before we use them in our model. Figure \ref{figure:KM} presents a plot of the Kaplan-Meier estimates of the survival probabilities. It is clear that the survival curve shows a long plateau that levels off to non-zero proportion, indicating the presence of a cured subgroup. As such, the proposed PCM-SVM model is appropriate here. We fit the PCM-SVM model and for comparison we also fit the PCM-Logit model. \textcolor{black}{ Since the sample size for the leukemia data is small, to deal with over-fitting or under-fitting, we adopt the 10-fold cross-validation technique that allows us to simultaneously fit and evaluate each model on the full data. This is consistent with the suggestion of Hastie et al. (2001).}

\begin{figure}[hptb!]
\centering
\begin{tabular}{cc}
		\includegraphics[scale=0.35]{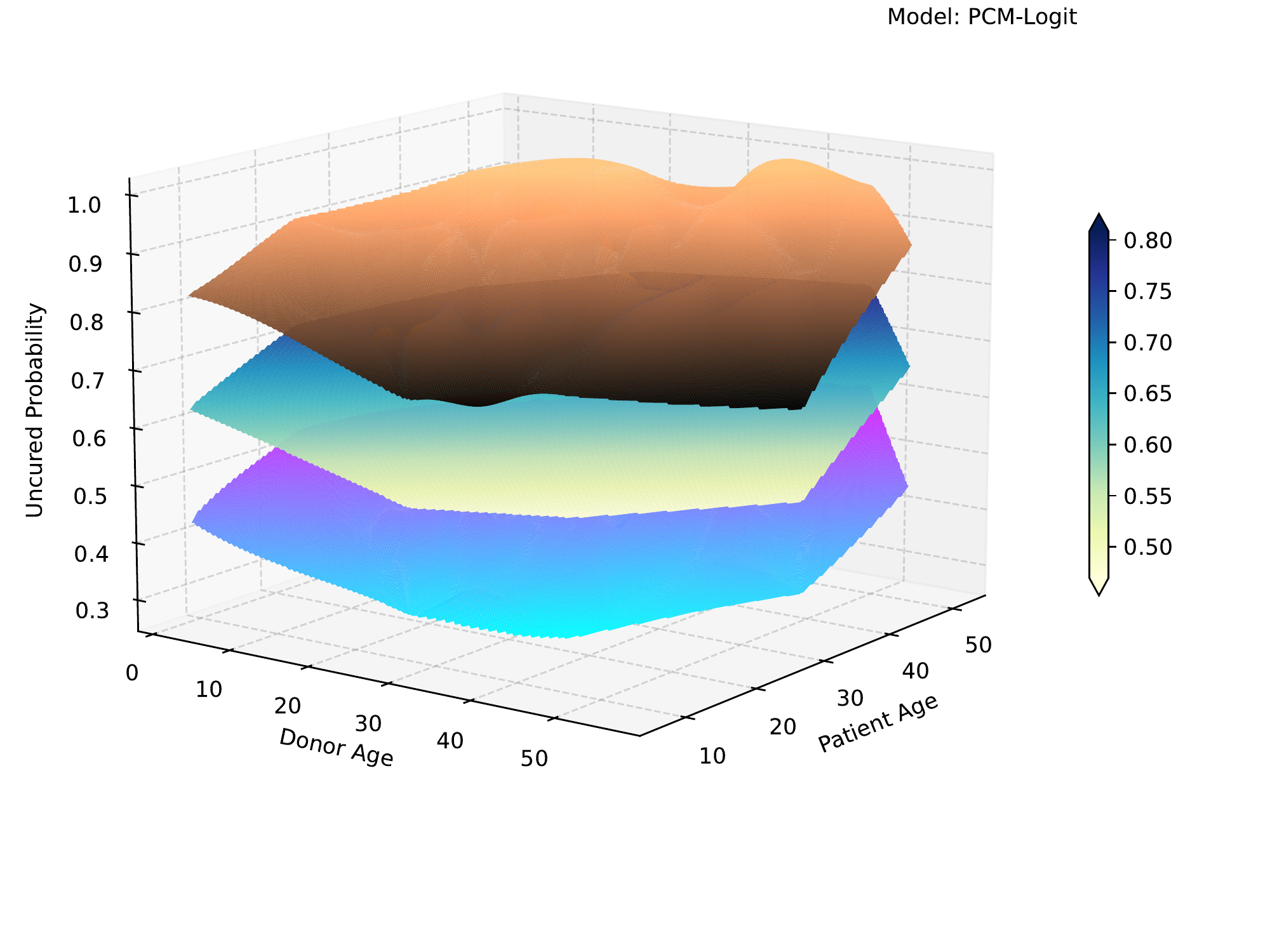}
		\includegraphics[scale=0.35]{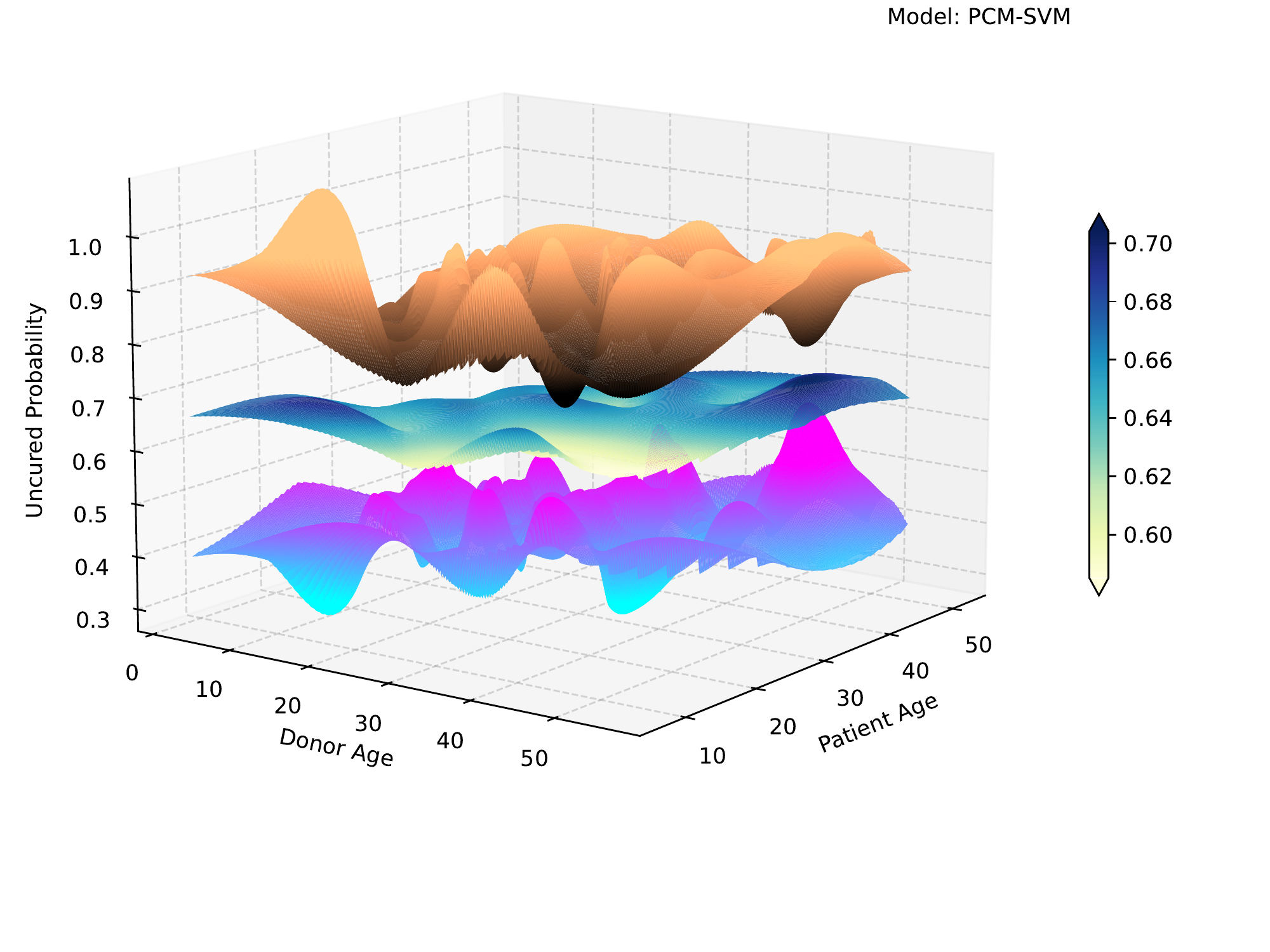}\\
		\includegraphics[scale=0.35]{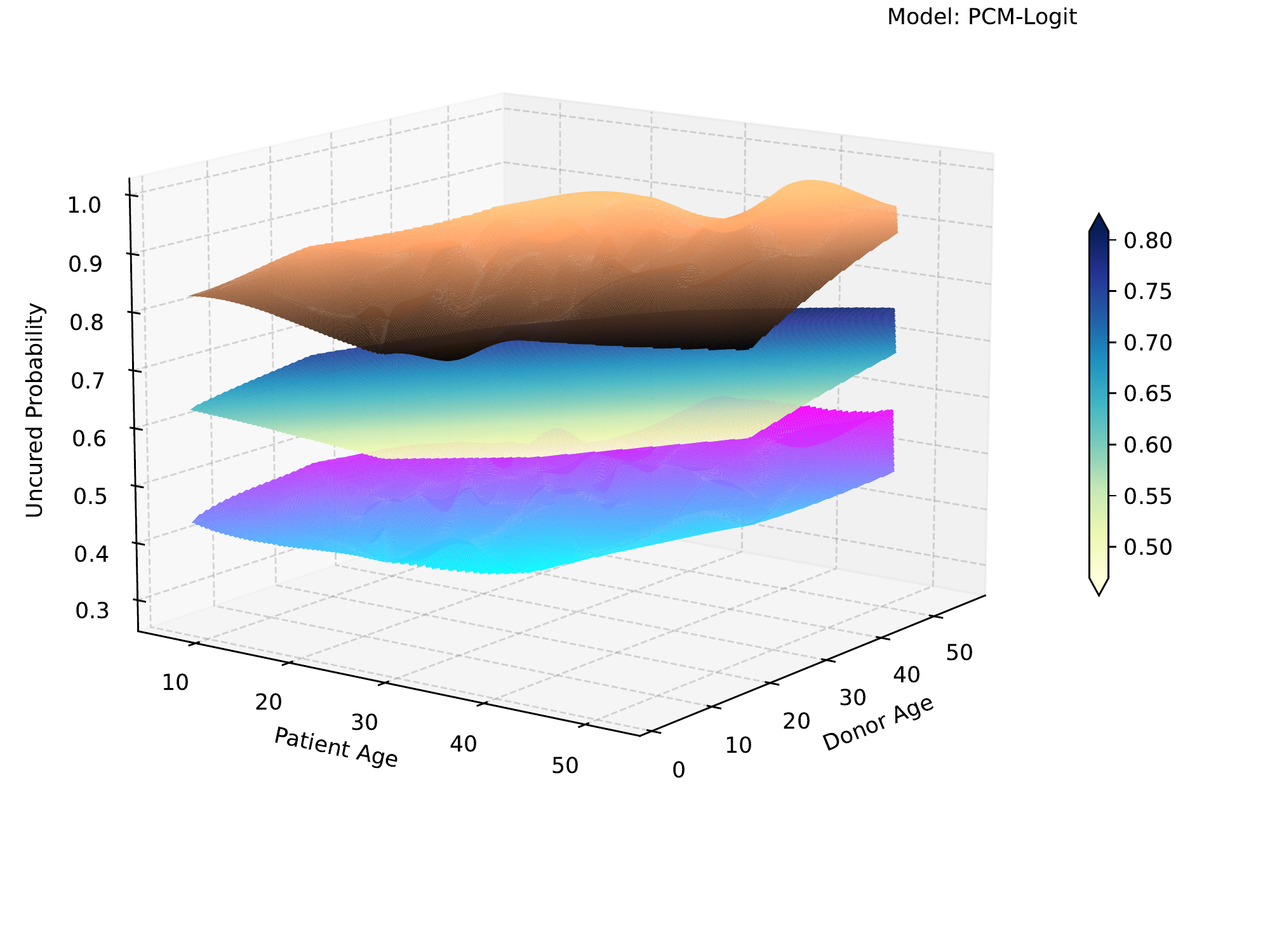}
		\includegraphics[scale=0.35]{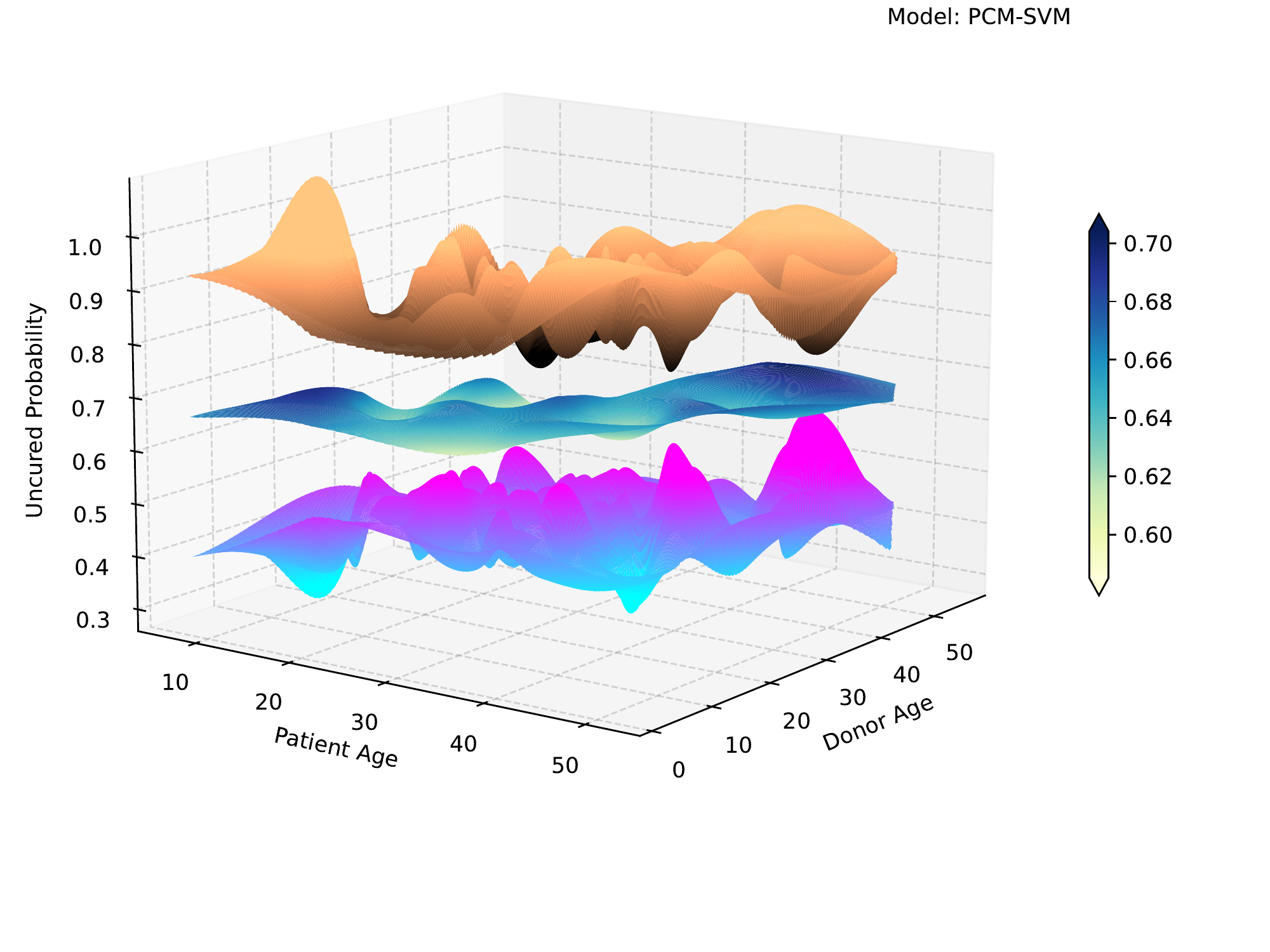}\\
\end{tabular}
  \caption{Plots of estimated non-cured probabilities as a function of donor's age and patient's age along with 95\% confidence bounds for the leukemia data.}
\label{figure:3D}
\end{figure}
\begin{figure}[htb!]
\centering
\includegraphics[scale=0.6]{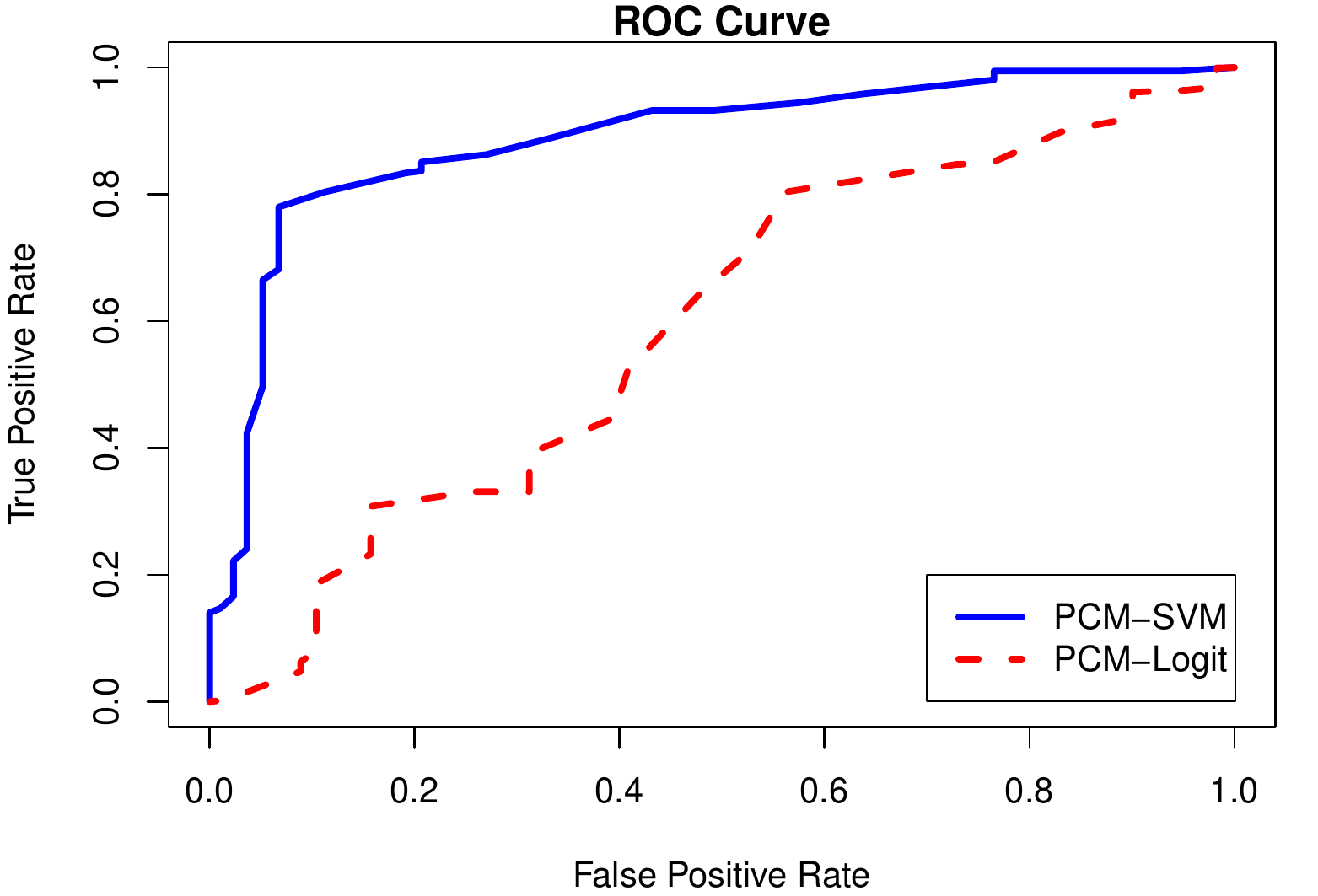}
\caption{ROC curves under different models for the Leukemia data}
\label{figure:ROC-BMT}
\end{figure}

First, we focus on the inference related to the incidence part of the PCM model. In Figure \ref{figure:3D}, we plot the estimates of the uncured probabilities, along with their 95\% confidence bounds, against patient's age and donor's age for both PCM-SVM and PCM-Logit models. Clearly, the PCM-SVM model can capture complex age effects on the uncured probability unlike the PCM-Logit model which can only capture monotonic relationships. To compare the predictive accuracies of the PCM-SVM and PCM-Logit models using the ROC curves and AUC values, we impute the missing cured statuses for the set of censored lifetimes. For this purpose, first, we estimate the conditional probability of non-cured by using eqn.\eqref{w} and the estimated values of $\pi(\bm x_i)$ and $F(t_i;\bm z_i)$ for $i=1,2,\cdots,n$. Then, using the estimated conditional probability of non-cured for each $i$, we generate a random variable from a Bernoulli distribution, which represents the cured/non-cured status. We repeat this process 500 times and present the averaged ROC curves in Figure \ref{figure:ROC-BMT}. The corresponding averaged AUC values (based on the testing set) for the PCM-SVM and PCM-Logit models turn out to be 0.887 and 0.592, respectively. Clearly, the proposed PCM-SVM model has a very satisfactory and much better predictive accuracy than the PCM-Logit model. \textcolor{black}{Note that our proposed simulation-based approach to estimate the missing cured statuses for the calculation of the ROC and AUC is preferred as it does not rely on naive assumptions such as the existence of a known ``cured time”, beyond which all censored observations are considered as cured (Asano et al., 2014). When compared to SVM-based mixture cure model of Li et al. (2020), the AUC value for Li et al.'s model turns out to be 0.713. Clearly, our proposed model again results in an improved predictive accuracy of cure.}


Next, we turn our attention to the inference related to the latency part of the PCM model. In Table \ref{table:MLE}, we present the estimates of the latency parameters together with their standard errors (SE) and $p$-values. The estimates of the standard errors are based on 100 bootstrap samples. The effects of patient's age and donor's age on the latency part turn out to be similar for both PCM-SVM and PCM-Logit models. However, there are differences when it comes to the significance of these effects. Since the estimate of $\beta_1$ is negative, it implies that the survival probability of a susceptible patient increases with an increase in the patient's age. However, such an effect is non-significant under the PCM-SVM model and significant at 10\% level of significance under the PCM-Logit model. On the other hand, since the estimate of $\beta_2$ is positive, it implies that with an increase in donor's age the instantaneous failure rate of a susceptible patient would increase. Such an effect is significant under both models at 10\% level of significance.

\begin{table}[htb!]
\color{black}
\center
\caption{\textcolor{black}{Estimation results corresponding to the latency parameters for the leukemia data}}
\begin{tabular}{lcc|cc|cc}
\hline
Parameter & \multicolumn{2}{c}{Estimates} & \multicolumn{2}{c}{SE} & \multicolumn{2}{c}{$p$-value} \\ \hline
&                            PCM-SVM & PCM-Logit& PCM-SVM& PCM-Logit & PCM-SVM  & PCM-Logit  \\ \hline
$\beta_1$ (Patient's Age) & -0.2241  & -0.6439  & 0.2818 & 0.4396 & 0.3264 & 0.0982 \\
$\beta_2$ (Donor's Age)   & 0.4321   & 0.7891   & 0.2832 & 0.4200 & 0.0877 & 0.0303 \\ \hline
\end{tabular}
\label{table:MLE}
\end{table}

\section{Conclusion and future work} \label{sect6}

With the rapid development in computing, machine learning techniques have become quite popular in medical applications (Floyd et al., 1994) as well as in traditional statistical modeling (Tandon et al., 2006) and standard survival analysis (Ishwaran et al., 2008; Spooner et al., 2020). However, its use in the context of cure rate model has only been recently investigated and not well explored. In the context of PCM model, there is only one recent study that has used neural network, a supervised machine learning algorithm, to incorporate unstructured covariates in the incidence part (Xie and Yu, 2021). To the best of our knowledge there are no other machine learning algorithms that have been used in the study of PCM model. The SVM is another well-known supervised machine learning algorithm that has received great attention in the study of a broad range of problems related to face detection, text categorization and pedestrian detection, among others. In the context of time-to-event data, the SVM has also been used in standard survival models (Fouodo et al., 2018; Van Belle et al., 2011) and, recently, in mixture cure rate model (Li et al., 2020). This manuscript is the first work that integrates the SVM with the PCM model. In particular, we have modeled the incidence part of the PCM model using the SVM to capture complex relationship between the covariates and the cured or non-cured probabilities, while preserving the simple interpretation for the latency part using the Cox's proportional hazards structure. For the estimation of model parameters, we have developed an EM algorithm and have made use of the sequential minimal optimization technique together with the Platt scaling method. \textcolor{black}{In the first part of our simulation study, we compared our proposed PCM-SVM with the commonly used PCM-Logit. The simulation results clearly show that the PCM-SVM can capture non-linearity in the data, which improves the accuracy and precision in the estimates of the uncured probabilities. Furthermore, we have shown that our model's ability to capture complex relationships also improves the estimation results corresponding to the latency part. Moreover, the PCM-SVM has better prediction and classification accuracies than the PCM-Logit, specifically when the true boundary separating the cured and non-cured subjects is non-linear. In the second part of our simulation study, we considered a more complicated scenario with 10 covariates (involving interaction terms and correlation) and compared the PCM-SVM with the PCM-Spline and PCM-NN. We showed that when it specifically comes to estimating the non-cured and overall survival probabilities, the PCM-SVM outperforms both PCM-Spline and PCM-NN. These findings clearly demonstrates the superiority of our proposed model. }

As an immediate followup work, it is of great interest to investigate the performance of the proposed model when the dimension of covariates is high, i.e., the sample size is less than the number of covariates. It is also of interest to develop goodness-of-fit tests to assess which of the two competing models (PCM-SVM and PCM-Logit) provide a better fit to a given data set. In case both models provide similar fits, it may be worth to use the existing PCM-Logit model so that the easy interpretation of the effects of covariates on the cured or non-cured probabilities can be preserved (Li et al., 2020). The PCM model assumes a Poisson distribution for the unobserved number of competing risks. Such an assumption does not allow simultaneous handling of over- and under-dispersion, which is commonly encountered in count data (Kokonendji et al., 2008). Thus, to provide more meaningful interpretation of results for data analysis, it is also of primary interest to explore the use of SVM to study flexible cure rate models that can simultaneously handle over- and under-dispersion (Pal, 2014; Pal and Balakrishnan, 2017; Pal, 2021). \textcolor{black}{In addition, the SVM as well as other machine learning techniques can be employed to extend the transformation family of cure models (Pal and Balakrishnan, 2017b; Wang and Pal, 2022) and the destructive cure models that look into the elimination of cancerous cells (Pal and Balakrishnan, 2016; Pal and Balakrishnan, 2017c,d; Pal et al., 2018).} We are currently working on these problems and hope to report the findings in future papers.

\begin{acks}[Acknowledgments]
The authors would like to thank the Associate Editor and three anonymous reviewers for their constructive comments which have led to a substantial improvement to an earlier version of this manuscript.
\end{acks}

\begin{supplement}
\stitle{Additional plots and R codes}
\sdescription{We provide additional plots corresponding to Section 4. We also provide R codes for data generation and SVM-based EM algorithm.}
\end{supplement}



\section*{References}

\begin{description}

\item Asano, J., Hirakawa, A. and Hamada, C. (2014). Assessing the prediction accuracy of cure in the {C}ox proportional hazards cure model: an application to breast cancer data. {\it Pharmaceutical Statistics}, {\bf 13}, 357--363.

\item Balakrishnan, N., Koutras, M. V., Milienos, F. S., and Pal, S. (2016). Piecewise linear approximations for cure rate models and associated inferential issues. {\it Methodology and Computing in Applied Probability}, {\bf 18}, 937--966.


\item Balakrishnan, N. and Pal, S. (2013). Lognormal lifetimes and likelihood-based inference for flexible cure rate models based on COM-Poisson family. {\it Computational Statistics \& Data Analysis}, {\bf 67}, 41--67.

\item Balakrishnan, N. and Pal, S. (2015a). Likelihood inference for flexible cure rate models with gamma lifetimes. {\it Communications in Statistics - Theory and Methods}, {\bf 44}, 4007--4048.

\item Balakrishnan, N. and Pal, S. (2015b). An EM algorithm for the estimation of parameters of a flexible cure rate model with generalized gamma lifetime and model discrimination using likelihood- and information-based methods. {\it Computational Statistics}, {\bf 30}, 151--189. 

\item Balakrishnan, N. and Pal, S. (2016). Expectation maximization-based likelihood inference for flexible cure rate models with Weibull lifetimes. {\it Statistical Methods in Medical Research}, {\bf 25}, 1535--1563.

\item Berkson, J. and Gage, R.P. (1952). Survival curve for cancer patients following treatment. {\it Journal of the American Statistical Association}, {\bf 47}, 501--515.

\item Boag, J.W. (1949). Maximum likelihood estimates of the proportion of patients cured by cancer therapy. {\it Journal of the Royal Statistical Society Series B}, {\bf 11}, 15--53.

\item Brown, E.R. and Ibrahim, J.G. (2003). Bayesian approaches to joint cure-rate and longitudinal models with applications to cancer vaccine trials. {\it Biometrics}, {\bf 59}, 686--693.

\item Cai, C., Zou, Y., Peng, Y. and Zhang, J. (2012). smcure: An R-package for estimating semiparametric mixture cure models. {\it Computer Methods and Programs in Biomedicine}, {\bf 108}, 1255--1260.

\item Chen, T. and Du, P. (2018). Promotion time cure rate model with nonparametric form
of covariate effects. {\it Statistics in Medicine}, {\bf 37}, 1625--1635.

\item Chen, M. H and Ibrahim J.G. (2001). Maximum likelihood methods for cure rate models with missing covariates. {\it Biometrics}, {\bf 57}, 43--52.

\item Chen, M.H., Ibrahim, J.G. and Sinha, D. (1999). A new Bayesian model for survival data with a surviving fraction. {\it Journal of the American Statistical Association}, {\bf 94}, 909--919.

\item Cooner, F., Banerjee, S. and Sinha, D. (2007). Flexible Cure Rate Modeling Under Latent Activation Schemes. {\it Journal
of the American Statistical Association}, {\bf 102}, 560--572.

\item Copelan, E. A., Biggs, J. C., Thompson, J. M., Crilley, P., Szer, J., Klein, J. P., Kapoor, N., Avalos, B. R., Cunningham, I. and Atkinson, K. (1991). Treatment for acute myelocytic leukemia with allogeneic bone marrow transplantation following preparation with BuCy2. {\it Blood}, {\bf 78}, 838--843.

\item Cortes, C. and Vapnik, V. (1995). Support-vector networks. {\it Machine Learning}, {\bf 20}, 273--297.

\item Floyd, C.E, Lo, J.Y., Yun, A.J., Sullivan, D.C. and Kornguth, P.J. (1994). Prediction of breast cancer malignancy using an artificial neural network. {\it Cancer}, {\bf 74}, 2944--2948.

\item Fouodo, C.J.K., Konig, I.R., Weihs, C., Ziegler, A. and Wright, M.N. (2018). Support vector machines for survival analysis with R. {\it The R Journal}, {\bf 10}, 412--423.

\item Hastie, T., Tibshirani, R. and Friedman, J. (2001). The Elements of Statistical Learning: Data Mining, Inference, and Prediction. New York: Springer.

\item Ishwaran, H., Kogalur, U.B., Blackstone, E.H. and Lauer, M.S. (2008). Random survival forests. {\it Annals of Applied Statistics}, {\bf 2}, 841--860.


\item Kokonendji, C.C., Mizere, D. and Balakrishnan, N. (2008). Connections of the Poisson weight function
to overdispersion and underdispersion. {\it Journal of Statistical Planning and Inference}, {\bf 138}, 287--1296.

\item Li, P., Peng, Y., Jiang, P. and Dong, Q. (2020). A support vector machine based semiparametric mixture cure model. {\it Computational Statistics}, {\bf 35}, 931--945.

\item Liu, H. and Shen, Y. (2009). A semiparametric regression cure model for interval-censored data. {\it Journal of the American Statistical Association}, {\bf 487}, 1168--1178.

\item Maller, R.A. and Zhou, X. (1996). {\it Survival Analysis with Long-Term Survivors}. John Wiley \& Sons Ltd.

\item Pal, S. (2014). Likelihood-based inferential methods for some flexible cure rate models. http://hdl.handle.net/11375/13688. Accessed 7 July 2021.

\item Pal, S. (2021). A simplified stochastic EM algorithm for cure rate model with negative binomial competing risks: An application to breast cancer data. {\it Statistics in Medicine}, {\bf 40}, 6387--6409.

\item Pal, S. and Aselisewine, W. (2023). Supplement to ``A semi-parametric promotion time cure model with support vector machine." 

\item Pal, S. and Balakrishnan, N. (2016). Destructive negative binomial cure rate model and EM-based likelihood inference under Weibull lifetime. {\it Statistics \& Probability Letters}, {\bf 116}, 9$-$20.

\item Pal, S. and Balakrishnan, N. (2017). Likelihood inference for COM-Poisson cure rate model with interval-censored data and Weibull lifetimes. {\it Statistical Methods in Medical Research}, {\bf 26}, 2093--2113.

\item Pal, S. and Balakrishnan, N. (2017b). Expectation maximization algorithm for Box-Cox transformation cure rate model and assessment of model misspecification under Weibull lifetimes. {\it IEEE Journal of Biomedical and Health Informatics}, {\bf 22}, 926$-$934.

\item Pal, S. and Balakrishnan, N. (2017c). An EM type estimation procedure for the destructive exponentially weighted Poisson regression cure model under generalized gamma lifetime. {\it Journal of Statistical Computation and Simulation}, {\bf 87}, 1107$-$1129. 

\item Pal, S. and Balakrishnan, N. (2017d). Likelihood inference for the destructive exponentially weighted Poisson cure rate model with Weibull lifetime and an application to melanoma data. {\it Computational Statistics}, {\bf 32}, 429$-$449.

\item Pal, S., Majakwara, J., and Balakrishnan, N. (2018). An EM algorithm for the destructive COM-Poisson regression cure rate model. {\it Metrika}, {\bf 81}, 143$-$171.

 \item Peng, Y. and Dear, K. B. (2000). A nonparametric mixture model for cure rate estimation. {\it Biometrics} , {\bf 56}, 237--243.

\item Peng, Y. and Yu, B. (2021). {\it Cure Models: Methods, Applications and Implementation}. Chapman and Hall/CRC.


\item Platt, J. (1999). Fast training of support vector machines using sequential minimal optimization. In B. Schlkopf, C. Burges, \& A. Smola (Eds.), {\it Advances in kernel methods - support vector learning} (pp. 185–208).  Cambridge, MA, USA: MIT Press.

\item Platt, J., et al. (1999). Probabilistic outputs for support vector machines and comparisons to
regularized likelihood methods. {\it Advances in Large Margin Classifiers}, {\bf 10}, 61--74.

\item Spooner, A., Chen, E., Sowmya, A., Sachdev, P., Kochan, N.A., Trollor, J. and Brodaty, H. (2020). A comparison of machine learning methods for survival analysis of high-dimensional clinical data for dementia prediction. {\it Scientific Reports}, {\bf 10}, 20410.

\item Tandon, R., Adak, S. and Kaye, J.A. (2006). Neural networks for longitudinal studies in Alzheimer’s disease. {\it Artificial Intelligence in Medicine}, {\bf 36}, 245--255.

\item Van Belle, V., Pelckmans, K., Van Huffel, S. and Suykens, J.A.K. (2011). Support vector methods for survival analysis: a comparison between ranking and regression approaches. {\it Artificial Intelligence in Medicine}, {\bf 53}, 107--118.

\item Wang, P. and Pal, S. (2022). A two‐way flexible generalized gamma transformation cure rate model. {\it Statistics in Medicine}, {\bf 41}, 2427--2447. 

\item Xie, Y. and Yu, Z. (2021). Promotion time cure rate model with a neural network estimated nonparametric component. {\it Statistics in Medicine}, {\bf 40}, 3516--3532.

\item Yakovlev, A.Y. and Tsodikov, A.D. (1996). {\it Stochastic Models of Tumor Latency and Their Biostatistical Applications}. World Scientific.

\item Yin, G. and Ibrahim, J.G. (2005). Cure rate models: a unified approach. {\it The Canadian Journal of Statistics}, {\bf 33}, 559--570.

\end{description}

\end{document}